\documentclass[aps,prc,twocolumn,showpacs,preprintnumbers,amsmath,amssymb,nofootinbib]{revtex4}
\usepackage[utf8]{inputenc}

\usepackage{color} %-- use \textcolor{color}{text} to write the text in color
\usepackage{ dsfont }
\usepackage{graphicx}

\usepackage{fancyhdr}
\usepackage{setspace}
\usepackage{amssymb}
\usepackage{relsize}
\usepackage{ tipa }
\usepackage{varwidth}
\usepackage{rotating}

\usepackage{pdflscape}
\usepackage{float}
\usepackage{hvfloat}

\def\pu1 {$^{241}$Pu$^*$\mbox{ }}

\newcommand{\etal}{{\it et al.}}

\begin{document}
\title{Questioning the $^{239}$Pu$(n,2n)^{238}$Pu cross section shape above  emission threshold}
\author{O.~Bouland}\email{olivier.bouland@cea.fr}
\affiliation{CEA, DES, IRESNE, DER, SPRC, Physics Studies Laboratory, Cadarache, F-13108 Saint-Paul-lez-Durance, France}
\author{V.~M\'eot and O.~Roig}
\affiliation{CEA, DAM, DIF, F-91297 Arpajon France\\
Université Paris-Saclay, CEA, LMCE, 91680, Bruyères le Chatel, France.}

\begin{abstract} 
%This paper reexamines the pattern of the $^{239}$Pu$(n,2n)^{238}$Pu reaction cross section from the threshold to the onset of third-chance fission at about 12~MeV.  
In the light of the JEF(F) European project longstanding story  according to the determination of the most exact shape of the $^{239}$Pu$(n,2n)^{238}$Pu reaction cross section and a recent measurement by M\'eot \etal, this paper aims to shed another light on this topic by bringing new theoretical feedback. To achieve this goal, the AVXSF-LNG computer program has been upgraded to model second-chance reactions using its decay-probability module and, then chained to the TALYS-ECIS06 nuclear reaction system of codes. Present diligent calculation of the (n,2n) cross section over the energy range from the threshold to the onset of third-chance fission at about 12~MeV, suggests that current evaluations under-estimate  the $^{239}$Pu(n,2n)  cross  section  below  10  MeV; under-estimation  of  the  order  of  7\%  relatively  to the JEFF-3.1 evaluation. On this ground, we propose an upward correction to the normalization of the measurement  by M\'eot \etal~ Correction factor of about 1.24 with a maximum uncertainty on present fitted model estimated to $\pm$11.6\%. Latter value is extracted from a sensitivity analysis of the calculation route to the level density model that is selected for the non-equilibrated residual nucleus and to  alternative choices we can make in terms of  neutron fission cross section measurement references for the  $^{238}$Pu  and $^{239}$Pu target nuclei.
\end{abstract}
\date{\today}

%\pacs{24.10.Pa,24.10.Lx,25.85.Ec}
%\keywords{} 

\maketitle
% tends to reach an asymptotic value (3-5\%)
%\section{\label{s:analysis}$R$-matrix theory analysis}
\section{\label{s:intro}Introduction}

A recent measurement of the $^{239}$Pu$(n,2n)^{238}$Pu reaction cross section right above threshold energy using the recoil method for counting $^{238}$Pu nuclei by M\'eot \etal~\cite{meo:21}, has  brought  a new piece of information according to the shape of the reaction. An important conclusion by the authors was the confirmation of the disagreement  between the set of measurements previously reported and the JEFF-3.3 evaluation~\cite{plo:20} in the vicinity of the $^{239}$Pu(n,2n) reaction threshold. A  list of the $^{239}$Pu(n,2n) published experiments is given in Table~\ref{tab:n2nexp}. Those measurements have been commented extensively by M\'eot \etal~\cite{meo:21} and McNabb \etal~\cite{mcn:01} including the experimental techniques used. Therefore we will not comment further in present paper except when actually needed.  
%Note EXP PAPER MEOT: The present paper confirms without ambiguity the disagreement observed between the previously published measurements and the JEFF3.3 evaluation in the region of the 239Pu(n, 2n) reaction threshold.
\begin{table}[t] 
\caption{\label{tab:n2nexp}. List of the measurements according to the $^{239}$Pu$(n,2n)^{238}$Pu reaction cross section.} 
\begin{center}
\footnotesize{
\resizebox{0.95\columnwidth}{!}{\begin{tabular}{|c|c|c|c|c|}
\hline
Authors&   \multicolumn {1}{c}{Range (MeV)} &\multicolumn {1}{c}{Method} &  \multicolumn {1}{c|}{Year} & Refs.\\
\hline
M\'eot \etal &   7-9  & Recoil & 2021  &~\cite{meo:21}\\
Becker \etal &   6-22  & Partial $\gamma$ raies & 2002  & ~\cite{ber:01,bec:02}\\
  Lougheed \etal &   13-15  & Activation & 2002 &~\cite{lou:02} \\
  Fr\'ehaut \etal &   6-13  & Direct  & 1980  &~\cite{fre:80}\\
 &    & neutron-counting &  &\\
 Mather \etal &   6-13  & Direct & 1972 &~\cite{mat:72} \\ 
  &    & neutron-counting &  &\\
%\etal &   6-22  & & +  \\
\hline 
 \end{tabular}}}
\end{center}
\end{table}

\begin{figure}[t]
\center{\vspace{0.cm}
\resizebox{0.95\columnwidth}{!}{
\includegraphics[height=5cm,angle=0]{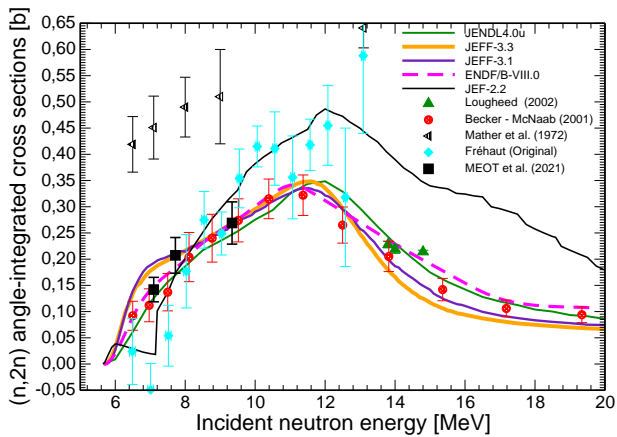}}}
\caption{(Color online) Comparison of the JEFF-3.3~\cite{plo:20}, ENDF/B-VIII.0~\cite{bro:18}, JENDL4.0u~\cite{jendl4u} and old JEF-2.2~\cite{for:97} evaluations (solid and dashed lines) with the experimental data (symbols) reported for the $^{239}$Pu(n,2n) reaction cross section. Experimental values plotted are those corrected by McNabb \etal~\cite{mcn:01} except the data by Fr\'ehaut \etal~and M\'eot \etal, which are respectively taken from EXFOR~\cite{exfor} and Ref.~\cite{meo:21}.}
\label{fig:n2nArt}
\end{figure}

The evaluation state-of-the-art according to the $^{239}$Pu(n,2n) angular-integrated reaction cross section  is well synthesized by Fig.~\ref{fig:n2nArt} on which are drawn from one side, experimental data and on the other side, some major evaluated curves. Recent evaluations (by contrast to the old JEF-2.2~\cite{for:97} data file) remain consistent except near threshold and at energies above the maximum of the excitation function. The JEFF-3.3 fast neutron energy evaluation (above 30~keV) has been performed by Romain \etal~\cite{rom:16} very carefully following the 'full model' methodology that enforces consistency over a unique nuclear data set according here to the Pu isotope series. The ENDF/B-VIII.0 $^{239}$Pu(n,2n)  evaluation~\cite{bro:18} is based on a rigorous least-squares analysis of a database limited to the measurements by Becker \etal~ and Lougheed \etal~  Indeed  the results of Fr\'ehaut \etal~ show much lower values around threshold than the selected database and exhibit very large uncertainties in the upper energy range. The oldest data set (1972) by Mather \etal~has been rejected by ENDF/B-VIII.0 because of full disagreement with all other data sets and poor experimental description.\\    

The $^{239}$Pu(n,2n)$^{238}$Pu reaction cross section has been extensively modeled over the years with various theoretical approaches~\cite{bla:84,cha:99,ros:99,che:00,rom:16,dup:17} or adjustment techniques~\cite{for:97,mcn:01,tom:08,bau:12}  but no definitive outcome was reached. Regarding the JEF(F) project, the question of the right magnitude and shape of the $^{239}$Pu(n,2n) cross section is a longstanding story that has been highlighted in 1997  at the time of the JEF-2.2 data file validation. Fort \etal~\cite{for:97} pointed out a severe conflict between the  (n,2n) excitation function shape as suggested by the Fr\'ehaut \etal~\cite{fre:80} measurement and the feedback brought by the analysis of an integral measurement  (the PROFIL program performed in the PHENIX reactor~\cite{tom:08}) that suggested, through a Bayesian adjustment based on the whole JEF-2.2 evaluated database,  a reduction factor of 1.55 to the Fr\'ehaut \etal~\cite{fre:80} data. However at the time of the evaluation, the preference was given to the experiment as acknowledged by the JEF-2.2 evaluated curve in Fig~\ref{fig:n2nArt}.\\  

According to present era, Fig.~\ref{fig:n2nArt} still carries issues about the most exact magnitude of the $^{239}$Pu(n,2n) cross section over the 7-10~MeV region, the correct shape at above 13 MeV and the ultimate disqualification of the experimental data by Fr\'ehaut \etal~\cite{fre:80}. Present paper is willing to consolidate the recent results by M\'eot \etal~\cite{meo:21} since the authors have emphasized that their measurements are still relative to the previously published data at high energy. Indeed, their uncertainties are dominated by the systematic uncertainty arising from the data of Becker \etal~\cite{ber:01,bec:02} adopted as a reference to normalize their data point at 9.34~MeV. The latter therefore carries a final uncertainty as large as 15\%, unfortunately propagated to the two  measured data points at 7.1 and 7.72~MeV. {\it The above  defines clearly the objective of present work}. First, to own the ability for assessing with a better precision the measurement normalization and possibly with the distance and hindsight, to address the most likely shape regarding the excitation function over the energy range up to its maximum.\\

We know that  at high energy where pure compound nucleus mechanism makes room for preequilibrium reactions, the way to compute the preequilibrium  plays a key role. For a long time modeled by classical exciton model of preequilibrium neutron emission~\cite{bla:83,kon:97}, quantum mechanical descriptions~\cite{fes:80,cha:07,dup:17} are now often privileged. In the TALYS software~\cite{kon:12} that we will be using in this work to compute the neutron flux feeding compound nucleus (CN) de-excitations, both classical (exciton) and quantum-mechanical (multi-step direct) descriptions are proposed.\\   

Present approach is willing to follow the footsteps of Romain \etal~\cite{rom:16} with the 'full model' methodology. This now conventional approach requires both efficient coupled channel optical model and preequilibrium process tools, together with a Hauser-Feshbach engine to split the total compound nucleus cross section among the open decay reaction compound channels. Although the TALYS program~\cite{kon:12} is able to manage easily the two sides of the calculation, present work has been using an independent Hauser-Feshbach algorithm made available across the AVXSF-LNG ({\it AVerage CROSS Section}  {\it F}ission - {\it L}ynn and {\it N}ext {\it G}eneration)  computer program; originally designed to analyse measured neutron-induced average cross sections of actinides in the statistical energy range from about 1 keV neutron energy up to the onset of second-chance fission~\cite{bou:13}. More recently extended to treat fission- and $\gamma$-decay-probability data~\cite{bou:19,bou:20} over the excitation range of the compound nucleus, this code has been here upgraded to model second-chance reactions using its decay-probability module. In present paper, our purpose is not to supply an exhaustive description of the latest version of the AVXSF-LNG code but rather to review the various features allowing reasonable modeling of cross sections up to third-chance fission. Latter developments associated to the best experimental and nuclear structure data, are expected to be the key stone of present evaluation task.\\ 

This article is organised as follows. Section~\ref{s:app} presents the computational approach and in particular how the TALYS software and the AVXSF-LNG code are chained. Section~\ref{s:HGt} recalls Hauser-Feshbach theory in particular dedicated to the fission decay channel in the R-matrix formalism both analytically and in a Monte Carlo way. Latter section details also the handling of second-chance reactions at above 5~MeV neutron energy and the way to model second-chance preequilibrium neutron emission using a macro-microscopic combinatorial method to simulate the level density of a pre-equilibrated compound system.  The next section~\ref{s:application} is devoted to present evaluation of the (\MakeLowercase{n}+$^{239}$P\MakeLowercase{u}) average cross sections over the [1~keV-12~MeV] energy range and in particular to the prediction of the (n,2n) total cross section. Finally, Section~\ref{ccl} presents our conclusion and recommendation about the shape and magnitude of the $^{239}$Pu(n,2n) cross section below the maximum of the excitation function with consequences on the normalization chosen in the recent measurement by M\'eot \etal

\begin{figure}[t]
\center{\vspace{1.cm}
\resizebox{0.9\columnwidth}{!}{
\includegraphics[height=5cm,angle=0]{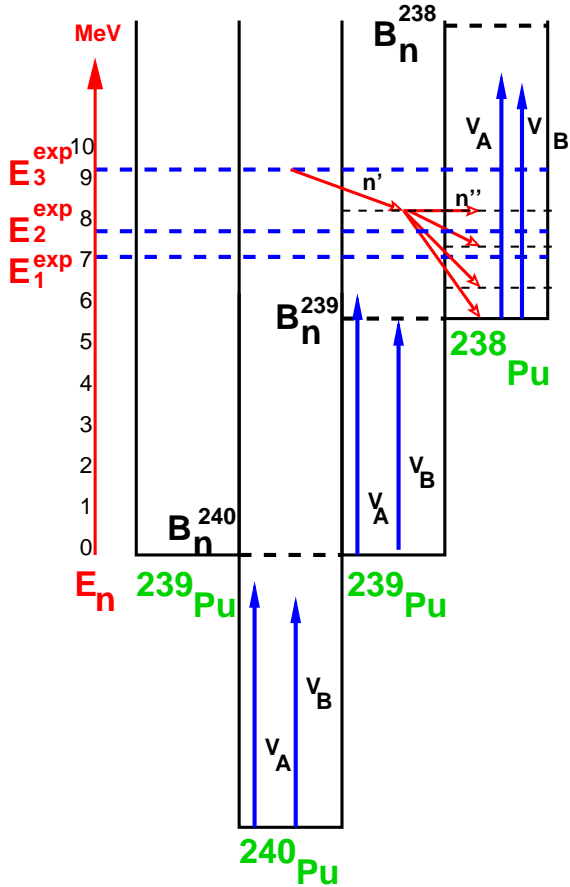}}}
\caption{(Color online) Schematic representation of nuclear potential wells for the $^{239}$Pu target/first-stage-residual/ second-stage-compound, $^{240}$Pu first-stage-compound and $^{238}$Pu second-stage-residual nuclei as involved in the (n,2n) two-stage process induced by neutrons for excitation energies at energies above  (B$_n^{240}$+B$_n^{239}$) in the $^{240}$Pu. Neutron beam energies corresponding to the recently measured (n,2n) cross section~\cite{meo:21}, are labeled E$^{\mbox{\footnotesize exp}}_{1,2,3}$. $V_A$ and $V_B$ mark the inner and outer fission saddle heights in the  compound systems involved.}
\label{potWells}
\end{figure}

\section{\label{s:app}Computational approach}

\subsection{\label{s:overall}Overall reaction channels picture}

\begin{figure}[t]
\center{\vspace{1.cm}
\resizebox{0.99\columnwidth}{!}{
\includegraphics[height=5cm,angle=0]{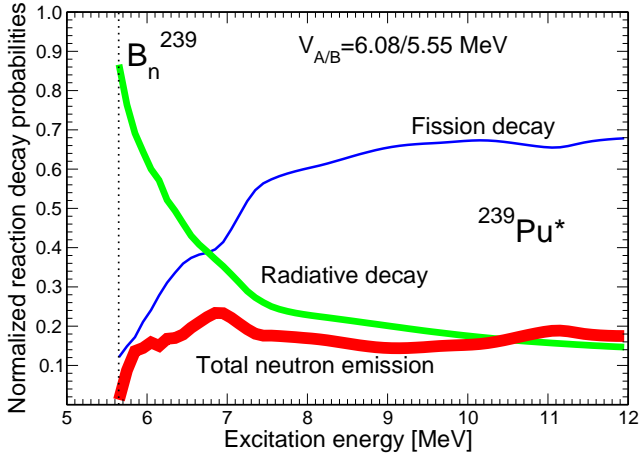}}}
\caption{(Color online) $^{239}$Pu compound nucleus reaction decay probabilities computed by Monte Carlo (below 7.2~MeV) or following pure  Hauser Feshbach path (above) as a function of  excitation energy. Thick, medium-thick and thin solid curves correspond respectively to total neutron emission-, $\gamma$- and fission-decay probabilities. Present spin-parity-integrated probabilities have been normalized to unity for display.}%
\label{fig:Pgfi239*}
\end{figure}

Before going further, it is useful to give an overview of the reactions involved in the calculation of the $^{239}$Pu$(n,2n)$ cross section. Assuming a $^{240}$Pu long-lived equilibrated system, meaning a CN, following a neutron capture by a $^{239}$Pu target  with neutron energies close to the $^{239}$Pu neutron binding energy (B$_n^{239}$ on Fig.~\ref{potWells}), second-chance reactions  occur. In this configuration,  high-energy excited levels in the $^{239}$Pu first-stage-residual nucleus reached by one-neutron emission, still carry enough excitation energy to decay by emitting a second neutron towards a second-stage-residual nucleus; namely the $^{238}$Pu. The latter will eventually de-excite by $\gamma$-emission; a reaction labeled (n,n'n''$\gamma_i$). In present case, we are not interested in reproducing a specific final ($i$) $\gamma$-ray emission channel, only in modeling the so-called total (n,2n)  reaction cross section. Figure~\ref{potWells} emphasizes well  the importance of carrying  best knowledge in terms of nuclear data for the $^{240}$Pu, $^{239}$Pu and $^{238}$Pu isotopes  and in particular accurate $^{239}$Pu decay probabilities to compute (n,2n), (n,n'$\gamma$) and (n,n'f)  reaction channels.  Latter spin-parity-integrated reaction probabilities are sketched in Fig.~\ref{fig:Pgfi239*}. We enlighten that second-chance fission is actually opened below B$_n^{239}$, depending upon  barrier height values (i.e.; on $V_A$ and $V_B$). However with a fission  probability magnitude lower than 14\%, the latter was not included  below B$_n^{239}$ in present calculation.            

\subsection{\label{s:over}Overall methodology}

Based on the coupled-channel iterative approach of the ECIS-06 code~\cite{ray:94,car:01} that extends the simple regime of the spherical optical model, present calculation using the TALYS program~\cite{kon:12} couples the five lowest lying levels of the rotational band~\cite{cap:09} built on the ground state of the target nucleus. Present calculation, as implemented in the  TALYS standard version, relies on the conventional approximation that assumes no correlation between direct and  CN processes to calculate the derived nuclear reaction cross sections. The flux going into the direct elastic and inelastic channels is subtracted from the total cross section to  calculate the compound nucleus formation cross section. It was demonstrated using  Engelbrecht-Weidenm\"uller (EW)  transformation by Kawano \etal~\cite{kaw:16} that the approximation of no correlation leads for actinides to  capture and fission channels corrections smaller than 2\% according to reference cross section calculations (report to Fig.~8 of Ref.~\cite{kaw:16}) over the [0-2]~MeV neutron fluctuating energy range. In the meantime the impact although larger, remains smaller than 7\% according to the inelastic channel and more specifically less than 2\% over the upper resonant region [1 to 2~MeV]. Those 2\% inaccuracy must be put in regards to corresponding experimental uncertainties that are much larger in general. As far as  present total (n,2n)  reaction cross section calculation is concerned, actual impact of the above approximation will be expected in the calculation of  second-chance neutron-emission probabilities for excitation energies lower than 1~MeV. This tough question will be reopened later in the paper. However by following common approach decoupling direct and CN processes, we remain consistent with the methodology carried until today in evaluation works and more specifically for JEFF-3.3.\\

\begin{figure}[t]
\center{\vspace{1.0cm}
\resizebox{0.99\columnwidth}{!}{
\includegraphics[height=5cm,angle=0]{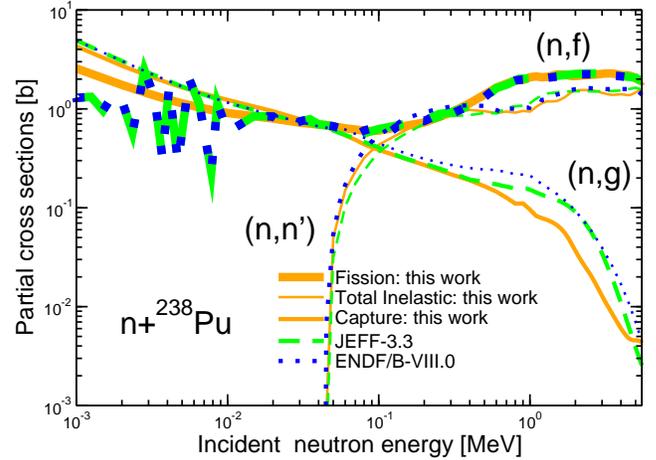}} }
\caption{\label{fig:pu239cnxs}(Color online) $^{238}$Pu neutron-induced partial cross sections (orange-solid curves) computed with AVXSF-LNG  and compared to some evaluated files (ENDF/B-VIII.0~\cite{bro:18}, JEFF-3.3~\cite{plo:20}). Thick, medium-thick and thin curves correspond  respectively to the $(n,f)$, $(n,\gamma)$ and $(n,n'_{tot})$ reactions.}
\end{figure}
\begin{figure}[t]
\center{\vspace{1.0cm}
\resizebox{0.99\columnwidth}{!}{
\includegraphics[height=5cm,angle=0]{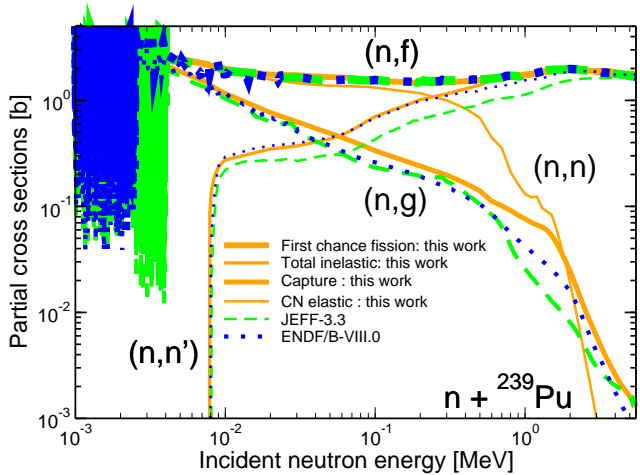}} }
\caption{\label{fig:pu240cnxs}(Color online) $^{239}$Pu neutron-induced partial cross sections (orange-solid curves) computed with AVXSF-LNG  and compared to some evaluated files (ENDF/B-VIII.0~\cite{bro:18}, JEFF-3.3~\cite{plo:20}). Thick, medium-thick, medium-thin and thin curves correspond  respectively to the $(n,f)$, $(n,\gamma)$, $(n,n'_{tot})$ and compound nucleus $(n,n)$ reactions.}
\end{figure}

Present strategy relies on a series of cross section adjustments reproducing available microscopic measurements (which average is well represented by the best evaluated cross section data) that lead to a unique nuclear database according to the (n+$^{238}$Pu) and (n+$^{239}$Pu) systems. The nice overall agreement between present adjusted cross sections  and current evaluated files is pictured on Figs.~(\ref{fig:pu239cnxs}) and ~(\ref{fig:pu240cnxs}) that span over the neutron energy range below second-chance fission. Our nuclear structure database is pretty much the same of the study~\cite{bou:13} where substantial effort has been put in building a suitable database of nuclear structure parameters. This was achieved across a simultaneous analysis of the neutron-induced average cross-sections of the plutonium isotopes from 236 to 244 for neutron energies in the range [1~keV - 5.5~MeV]. However in the study~\cite{bou:13} no rigorous connection was made between the observed total cross section and the total compound nucleus cross section feeding the Hauser-Feshbach (HF) equations~\cite{hau:52}. This point, potentially  fragile when investigating higher neutron energies, has been clarified with corresponding TALYS-ECIS06 calculations as input to present study as it will be summarized in the following.\\

Following the TALYS-ECIS06-based methodology, the total cross section is the simple sum of $\sigma^{direct-el}$, the direct elastic cross section (i.e., the shape elastic) and $\sigma^{R}$, the reaction cross section. In the absence of EW transformation, the total elastic cross section can be calculated as the incoherent sum of direct and CN components meaning, 
\begin{eqnarray}
\sigma_{nn,tot} = \sigma_{nn}^{direct-el}   + \sigma_{nn}^{CN}         \mbox{  }\mbox{  }\mbox{  }\mbox{  ,}
\label{eq:totel}
\end{eqnarray}
where the incident neutron energy dependence, $E_n$, has been left aside for display (as in the next equations). On the same footing, the total inelastic cross section is
\begin{eqnarray}
\sigma_{nn',tot} = \sigma_{nn'}^{direct-inel} + \sigma_{nn'}^{CN}     \mbox{  }\mbox{  }\mbox{  }\mbox{  ,}
\label{eq:totinel}
\end{eqnarray}
where the direct inelastic cross section, $\sigma_{nn'}^{direct-inel}$, is the sum of three components meaning of, 
\begin{enumerate}
\item the discrete states explicitly treated across the coupled channel optical model  (CCOM) equations,
\item the weakly-coupled discrete states of somewhat higher energy which  estimation is based on the distorted wave Born approximation~\cite{kon:12} and finally,
\item the excited states within the continuum that are far from being equilibrated. Associated calculation starts at 1~MeV neutron energy in present calculation.
\end{enumerate}
Latter component shape and magnitude depends strongly on the selected preequilibrium (PE) model. The Equation~\ref{eq:totinel} can be reformulated as, 
\begin{eqnarray}
\sigma_{nn',tot} = \sigma_{nn'}^{direct-disc.} + \sigma_{nn'}^{PE} + \sigma_{nn'}^{CN}     \mbox{  }\mbox{  }\mbox{  }\mbox{  ,}
\label{eq:totinel1}
\end{eqnarray}
in which common assumption is made that direct discrete, preequilibrium and CN inelastic cross sections can be added incoherently. Subsequently the fraction of remaining flux (also represented by the {\it composite-nucleus formation} cross section according to TALYS terminology), $\sigma^{CF}$, and feeding the preequilibrium stage is,  
\begin{eqnarray}
\sigma^{CF} = \sigma^{R} - \sigma_{nn'}^{direct-disc.}     \mbox{  }\mbox{  }\mbox{  }\mbox{  .}
\label{eq:cf}
\end{eqnarray}
%~\cite{hau:52}

In this paper, we will not discuss much about the question of the preequilibrium process that has been a long-term research topic for decades since at least~\cite{gri:66}. As previously mentioned, the  TALYS standard version offers two  alternatives for the preequilibrium calculation: the first, based on a two-component exciton (particle-hole pair) classical model~\cite{kon:12}, is recommended with the assumption of  a spin distribution for the preequilibrium process close to a compound nucleus spin distribution (TALYS default option) and the second referring to multistep direct (MSD)-multistep compound (MSC) quantum mechanical preequilibrium theory~\cite{kon:97}. Literature on the exciton model reports lack of accuracy to describe angular distributions whereas the MSD/MSC  tends to reproduce measured angle-integrated emission spectra with an accuracy close to that found in the classical model. A sensitivity study to the preequilibrium model is proposed in the Appendix to quantify its impact on the calculated (n,2n) cross section.  Next step in this paper  will then be the  description of the compound nucleus decay process as managed by the AVXSF-LNG code. Before that, we might portray the actual chain of collisions to be modeled.\\

When a bombarding low-energy neutron enters a target nucleus and collides with a nucleon, it is very unlikely that either of the particles emerging from that initial collision will have enough energy to escape from the nucleus and it will go on to have further collisions and eventually the system equilibrates, i.e. being a compound nucleus. When the bombarding neutron energy is increasing,  a neutron may escape after the initial neutron-nucleon collision  (the preequilibrium phase). That initial collision can no longer result in further internal collisions and hence the associated flux must be removed from the next-stage CN formation. This stage is modeled, similarly to TALYS, by the introduction of an {\it escape factor} that governs the compound nucleus formation cross section, $\sigma^{CN}$.  The actual flux feeding in this work the compound nucleus decay process is finally obtained by the following  (neutron energy-dependent) equivalence, 
\begin{eqnarray}
&&\sigma^{CN} =  \sigma^{CF}   - \sigma_{nn'}^{PE}   = \sigma^{CF}\biggl[1-\frac{\sigma_{nn'}^{PE}}{\sigma^{CF}}\biggr] \mbox{  }\mbox{  }\mbox{.}
\label{eq:totcn}
\end{eqnarray}
where the ratio $\biggl[\frac{\sigma_{nn'}^{PE}}{\sigma^{CF}}\biggr]$, defines the preequilibrium phase escape factor. 

The neutron energy-dependent preequilibrium cross section as an input to present study, is supplied by the TALYS code once the CCOM calculations have been performed by ECIS-06 as well as the calculation of the {\it direct-cross-section-eliminated} transmission coefficients (Ref.~\cite{kaw:16}; see also Appendix~\ref{connect})  needed to compute the preequilibrium cross section. Finally those transmission coefficients (or equivalently the closely-related optical model strength functions) are used to solve the HF equations. %See Talys 1.95 manual page 85 
Next paper section sketches the HF formalism with special emphasis on the fission channel treatment as implemented in our computer program. For completeness, the relationship between the entrance optical model strength functions, as newly input to the AVXSF-LNG code, and the compound nucleus formation cross section is reported in the Appendix~\ref{connect}.  

\section{\label{s:HGt} Hauser-Feshbach theory}

The  partial average cross section $\sigma_{cc'}$ formulation for an entrance channel $c$ and exit channel $c'$  applied to neutron-induced reactions for given neutron energy $E_n$, is adequately described by Hauser-Feshbach statistical theory~\cite{hau:52} with $W_{c,c'}$, the in-out-going channel width fluctuation correction factor~\cite{mol:61} (named in the following {\it customary WFCF}). This reads   
\begin{eqnarray}
{\sigma_{n,c'}}(E_n)&=& \mathlarger{\sum_{J^\pi}}  \Biggl[ \sigma_{n}^{CN}(E_n,J^{\pi})  \nonumber \\
\times  \sum _{s'={|I'-i'|}}^ {{I'+i'}}&& \sum _{l'={|J-s'|}}^ {J+s'}  {\frac{T_{c'}^{J^{\pi_{(l's')}}} (E_{c'})}  {\sum_{c''} {T_{c''}^{J^{\pi_{(l"s")}}} (E_{c''})}}}\times W_{n,c'}^{J^\pi} \Biggr]\mbox{ , }\mbox{ }\mbox{ }\mbox{ }\mbox{ }
\label{eq:hfwccp}
\end{eqnarray}
where $\sigma_{n}^{CN}(E_n,J,{\pi})$ is the neutron-induced partial compound nucleus formation cross section related to  a given ($J,\pi$) couple; the expression of which is supported by Eq.~\ref{eq:NC}. Equation~\ref{eq:hfwccp} can be written concisely as,
\begin{eqnarray}
{\sigma_{n,c'}}(E_n)=\sum_{J^\pi}  \Biggl[ \sigma_{n}^{CN}(E_n,J^{\pi})\times  \mathcal{P}_{c'}^{J^\pi} (E_{c'})\times W_{n,c'}^{J^\pi}\Biggr]
\label{eq:hfhfwconcice} 
\end{eqnarray}
where $\mathcal{P}_{c'}^{J^\pi} (E_{c'})$ is the individual decay probability into channel $c'$ from given $(J,\pi)$ state at the corresponding CN excitation energy $E_{c'}$. 
The $W_{c,c'}$ factor is especially important because this is the recognition of the role played by statistical fluctuation effects in the CN states governing the transmission coefficients and also the interference effects of these states. $W_{cc'}$ is calculated by numerical integration of the general one dimension integral given by Dresner~\cite{Dre:57} assuming a non-fluctuating capture width and, for other reactions, a $\chi^2$  width distribution with $\nu$ effective degrees of freedom for $n$ open reaction channels.  Channel width fluctuations from resonance to resonance have an important effect on average cross sections. When only a few channels are open, the $W_{cc'}$  terms will significantly reduce the value of partial reaction cross sections (e.g; as large as -40$\%$ in the case of the inelastic reaction for the (n+$^{239}$Pu) system - Fig.8 of Ref.~\cite{bou:19}). This contrasts with the elastic scattering channel that is enhanced by up to a value of 2.5 for the same system.  The overall effect decreases with increasing number of channels and thus with excitation energy. For actinides, this correction becomes negligible 1.6 MeV above $B_n$; statement also made in comparison with other sources of uncertainties in the final evaluation of the transmission coefficients.\\ 

Beyond the elastic and inelastic neutron channel transmission coefficients to be calculated following Eq.~\ref{eq:tn}, the capture and fission transmission coefficients rely on  classic narrow resonance approximation that is valid at the limit of small strength functions. It reads,
%In terms of radiative average cross sections ($\sigma_{n,\gamma}$), significant differences show up in between the various curves above several hundred of keV where no experimental data are available and where the very small value of the capture cross section complicates any  measurement. Present $\sigma_{n,\gamma}$ calculation rely for the $\gamma$-decay Hauser-Feshbach transmission coefficients, $T_\gamma^{J^{\pi}}$, on the general form established by Moldauer~\cite{mol:67} at the limit of very small strength function, that is 
\begin{eqnarray}
T_\gamma^{J^{\pi}} =   2\pi\frac{\bar\Gamma^{J^{\pi}}_{\gamma}}{D_{J^{\pi}}}.
\end{eqnarray}
where $D_{J^{\pi}}$ is the mean average resonance spacing for given spin $J$ and parity ${\pi}$ and   $\bar\Gamma^{J^{\pi}}_{\gamma}$, the corresponding  total radiative  capture average width. Prior estimates of $\bar\Gamma^{J^{\pi}}_{\gamma}$  for each Pu isotope were obtained from theoretical considerations supplemented by the  use of a modified  version of the Kopecky-Uhl model~\cite{kop:90} for  the radiative strength function ($S_\gamma$). Since the total radiation widths calculated in this way broadly agree no more than 10 to 20\% with measured radiation widths, we have made a further empirical adjustment to them so that for $^{240}$Pu+n the calculation becomes in very close agreement with the radiation width measurement by J.A.~Harvey \etal~\cite{har:82} at the 1.06 eV $s$-wave resonance ($\Gamma_{\gamma}=30.7\pm0.6$ meV). The latter appears to be the most accurate radiation width measurement available in the literature. Best estimated values used at neutron separation energy ($B_n$) according to $s$-waves are 39.0 and 30.6~meV respectively for the $^{240}$Pu and $^{241}$Pu compound systems. As a reference, corresponding values in the Atlas of Neutron resonances~\cite{mug:06} are (43$\pm$4)  and (31$\pm$2)~meV. We have also used this model to calculate the energy dependence of the radiation widths. The latter exhibit a general increase in energy, which we have expressed approximately by applying an exponential factor to the value calculated at the binding energy, the 'temperature' parameter in this factor being of the order of 3 MeV. This energy variation of the radiation width affects considerably the value of the radiative capture cross-section at the higher neutron energies. Finally, we have taken into account the reduction of the capture cross-section by the lower energy primary gamma rays resulting in $(n,\gamma n')$ and $(n,\gamma f)$ branching.~\cite{lyn:18}

\subsection{\label{ss:HFf}Hauser-Feshbach formulation for fission}

Equation~\ref{eq:hfhfwconcice} applied to the fission channel becomes

\begin{eqnarray}
{\sigma_{n,f}}(E_n)=\sum_{J^\pi}  \Biggl[ \sigma_{n}^{CN}(E_n,J^{\pi})\times  \mathcal{P}_{f}^{J^\pi} (E_{f})\times W_{n,f}^{J^\pi}\Biggr]\mbox{ . }
\label{eq:hfhfwconcicef}
\end{eqnarray}
In the peculiar case of the fission channel, $W_{nf}$, the exit fission channel must be seen as an overall fission channel across a double-humped barrier (possibly triple-humped) with for the associated  average fission width statistical distribution a corresponding number of degrees of freedom (DoF), $\nu_f$, altered by the intermediate structure effect~\cite{bou:13}. This DoF now so-called effective, $\nu_f^{eff}$, holds a value~\cite{bou:14} between zero and unity, that invalidates the common hypothesis $\nu_f=1$. Taking into account of the existence of at least a secondary well on the path to fission with associated second well eigenstates (meaning the class-II states) properties, the Hauser-Feshbach average cross section according to the fission reaction is best modeled  as,
\begin{eqnarray}
&& \sigma_{nf}(E_n) = \mathlarger{\sum_{J^\pi}} \Biggl[ \sigma_{n}^{CN}(E_n,J^{\pi}) \times \nonumber \\
&&\Big[ \mathsmaller{\sum_{\mu\in J^\pi}}  \mathcal{P}_f(E_{f},\mu)  W_{II}(\mu) \Bigr] W_{n,f}^{J^\pi}(\nu_f^{eff}) \Biggr]  \mbox{ , }  
\label{eq:FXSdecoupled} 
\end{eqnarray}
with $W_{II}(\mu)$, the {\it class-II state width fluctuation correction factor}, associated to a specific outer barrier transition state ($\mu$). This includes the concept of transition state proposed by A.~Bohr~\cite{Boh:56} who showed that important post-scission behavior aspects of the fission products, such as overall angular distribution, are dependent on the state of intrinsic excitation at the saddle point and suggested that these intrinsic states (also referred to transition states) should be regarded as the {\it physical} channels. This statement is important to be noticed since for a strict reaction theory, such as R-matrix theory, fission comprises a very large number of channels, which are formally defined at or beyond the scission point as the incipient fission product pairs in a multitude of states of excitation energy and angular momentum relationships. The $J^\pi$-integrated patterns of the two fluctuating factors, $W_{II}$ and $W_{n,f}$, still calculated by numerical integration of  the Dresner general  one dimension integral, have similar magnitude in the case of the $^{239}$Pu target fissile nuclide with a reduction of the average fission cross section in between 10 to 20 $\%$ each.\cite{bou:14} Latter  modeling is implemented in our code~\cite{bou:13}. Whenever the standard Eq.~\ref{eq:hfhfwconcicef} is used, we expect some parameters compensation during the adjustment. The Equation (\ref{eq:FXSdecoupled}) does not provide any guidance on the way to calculate the fission probability, $\mathcal{P}_f(E_{f},\mu)$. The formulation of the latter depends clearly on the strength of the coupling between class-I and class-II states which is a function of barrier heights and curvatures as well as the excitation energy in the compound nucleus. Approximate formulae valid in specific coupling situations are implemented in the  code and give good estimates of the average fission cross section. However a Monte Carlo  option is used when analytic approximations fail, a full sampling of the level characteristics (average spacings and widths) distributions is then needed. Next paragraph gives some hints and lists some references about the Monte Carlo implementation in the code; keeping in mind that this time-consuming procedure provides an added value only in the energy region where  cross sections or probabilities actually fluctuate.

\subsection{\label{ss:HFMC}Calculation of average cross sections \\ in a Monte Carlo (MC) framework} 

An efficient alternative to analytical expressions~\cite{bjo:80} valid only under specific conditions is the Monte Carlo-type method. The MC procedure presents the advantage of providing average cross sections taking full account of  statistical nuclear data parameter fluctuations  under the relevant intermediate structure coupling condition.  Our present approach simulates $R$-matrix resonance properties of a selected class-II state and those of the overlapped neighboring class-I states, over at least a full class-II energy spacing, using a chain of pseudo-random numbers for a fine-tuned selection process based on both level width and spacing statistical distributions with suitable averages. Most of the components of Eq.~\ref{eq:FXSdecoupled} become closely tied  and cannot be separated anymore. One realizes immediately the high potential of the MC procedure that carries a compact formulation of the pre-cited equations. In terms of the average fission cross section formulation it follows,
\begin{eqnarray}
{\sigma_{nf}}(E_n)=\mathlarger{\sum_{J^\pi}}  \Biggl[\sigma_{n}^{CN}(E_n,J^{\pi}) \mathcal{P}_{f;MC-n}^{J^\pi} (E_{f})\Biggr]\mbox{ ,}
\label{eq:xsmc}
\end{eqnarray}
with $\mathcal{P}_{f,MC-n}^{J^\pi}$, the exact average neutron-induced fission probability for given  ($J,\pi$) computed by MC treatment. In our work, the dual analytical and MC approaches have been followed for calculating the most accurate average cross-sections for each partial wave versus excitation energy. When the number of open channels becomes large, the approximated formulae are accurate enough to determine the fluctuation averaging factors. We emphasize that the correct modeling of the intermediate structure intermixed with level parameters fluctuations can change the estimate of barrier heights by a few  hundreds of keV from values deduced using the standard HF formulation (Eq.~(\ref{eq:hfhfwconcice})). The present MC technique extensively described in Refs.~\cite{bou:13,bou:14} will not be commented further.

\subsection{Treatment of second-chance reactions}

\subsubsection{General compound nucleus  picture}

\begin{figure}[t]
\center{\vspace{1.cm}
\resizebox{0.99\columnwidth}{!}{
\includegraphics[height=5cm,angle=0]{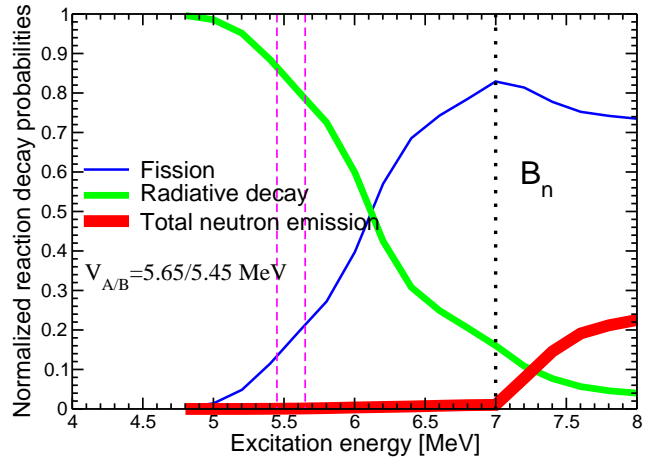}}}
\caption{(Color online) $^{238}$Pu CN reaction decay probabilities (computed by Monte Carlo below 7.5~MeV) as a function of the excitation energy. Thick, medium-thick and thin solid curves correspond respectively to total neutron emission, $\gamma$ and fission decay probabilities. Present spin-parity-integrated probabilities have been normalized to unity for display.}%
\label{fig:Pgfi238*}
\end{figure}

As introduced, the AVXSF-LNG code was originally designed to analyse measured neutron-induced cross sections from the upper end of the energy-resolved resonances range to the onset of second-chance fission. modeling of higher energy cross sections  require additional developments since two-step reactions of another kind, by reference to the $(n,\gamma n')$ and $(n,\gamma f)$ reactions  mentioned above that involve the same compound nucleus, may occur. It happens in the framework of a residual nucleus acting as a second-stage compound nucleus. This configuration is verified  when the decay of the compound nucleus goes by neutron emission with a residual nucleus left in a state of excitation energy close to the neutron separation energy in this same  residual nucleus. That kind of second-chance reactions is also described as {\it multiple compound nucleus emission} when treated in the framework of the HF equations.  Regarding  present study of the (n+$^{239}$Pu) interaction from 1~keV up to 12~MeV, this can be depicted by the following schematic diagram,
\begin{eqnarray}
&&\mbox{n''}\mbox{ + }^{238}\mbox{Pu}^{*}\nonumber \\
\nearrow &&\nonumber \\
\mbox{n + }^{239}\mbox{Pu}\rightarrow\mbox{ }^{240}\mbox{Pu}^*\rightarrow\mbox{ n'}\mbox{ +}\Bigl[^{239}\mbox{Pu}^*\Bigr]\rightarrow&&\gamma_i + ^{239}\mbox{Pu}^{*} \nonumber\\%Pu*$^{239}$Pu+n'$         (n,2n)$^{238}$Pu   $n$+(A-1) A$^*$ 
\searrow&&\nonumber\\
&&\mbox{Fission}\label{eq:symbol}
\end{eqnarray}
where the 'star' characterizes the nucleus in any state of excitation (including the ground state of zero-excitation energy).  
We realize promptly that  treatment of  second-chance reactions involves proper knowledge of the nuclear properties of both the first- and second-stage-residual nuclei (Fig.~\ref{potWells}). Present work has been eased by a previous cross section analysis of the plutonium isotope series made over the lower energy range [1~keV - 5.5~MeV]~\cite{bou:13}. In particular the $^{239}$Pu decay probabilities (Fig.~\ref{fig:Pgfi239*}), for which the knowledge is requested to compute (n,2n), (n,n'$\gamma$) and (n,n'f) two-step reactions, rely mostly on the average parameters, level densities and barrier parameters evaluated previously. %Details on the approach taken here in terms of probability normalization and choice in the correlations among the various channels, will be acknowledged later in the text.       
%Barrier parameters are usually derived from experimental observations and theoretical assessments and so,
Table~\ref{tab:barriers} displays the set of evaluated fission barrier heights and curvatures parameters used in this work. A larger comparison of parameters against reference data derived from experimental observations or theoretical assessments, is shown in Ref.~\cite{bou:13} (report to Figs.~(8) and (9)). One notices from Tab.~\ref{tab:barriers}, the strong fissile character of the $^{238}$Pu compound system with barrier heights much lower than corresponding neutron emission threshold and a subsequent fission probability magnitude reaching 0.8 at $B_n$ (Fig.~\ref{fig:Pgfi238*}). Thus we expect a competition start between the (n,2n) and  (n,2nf) channels  at a  neutron energy significantly lower than the one between the  (n,2n) and (n,3n) channels since the later is strictly conditioned to the value of $B_{n}^{238}$.  

%-- TABLE -- Fission Barrier Parameters
\begin{table}[ht]\center{
\caption{\label{tab:barriers}Fission barrier parameters fine-tuned in this work for Pu isotope cross section modeling. Neutron binding energies ($B_n$) quoted enlighten the fissile or fertile nature of each isotope. For comparison, the HFB-BCS predictions by Goriely  \etal~\cite{gor:09} available from RIPL-3~\cite{cap:09} are listed as well as those of Ref.~\cite{bou:13}.}
\footnotesize{\resizebox{0.9\columnwidth}{!}{\begin{tabular}{cc|cccc|c}
\hline\hline
Fissioning &  & $V_A$ & $\hbar \omega_A$ & $V_B$ & $\hbar \omega_B$ & $B_n$\\
Pu isotope  &&  \multicolumn{4}{c|}{[MeV]}&[MeV]\\
% & & & & & \\
\hline
& This work &{\bf  5.65 }& {\bf 1.05 }& {\bf 5.45 }& {\bf 0.60 } & 7.00\\
238 & Ref.~\cite{bou:13} & 5.65 & 1.05 & 5.45 & 0.60  & \\
& HFB-BCS & 5.96&	0.67&	5.24&	0.39\\
\hline
&   This work & {\bf 6.08 }& {\bf 0.98 }& {\bf 5.55 }& {\bf 0.55 } & 5.65\\
239 &  Ref.~\cite{bou:13} & 6.05 & 0.98 & 5.55 & 0.55  & \\
& HFB-BCS &5.96	&0.67	&5.33&	0.42 \\
\hline
& This work & {\bf 5.65 }& {\bf 1.05 }& {\bf 5.33} & {\bf 0.60 } & 6.53\\
240 & Ref.~\cite{bou:13}  & 5.65 & 1.05 & 5.23 & 0.60  & \\
& HFB-BCS & 6.49 &	0.70&	5.61&	0.37\\
\hline \hline
\end{tabular}}}}
\end{table}

\subsubsection{Derived equations}

Since in the neutron continuum energy range when $E_n>2.0$~MeV the pure Hauser-Feshbach  equation (Eq.~(\ref{eq:hfwccp}) with WFCF set to 1) becomes valid, the two-neutron-emission spin-parity partial cross section with second-stage-residual nucleus left in a low-lying state of excitation  $k$, can be calculated as,
%, the so-called (n,2n) cross section of the $^{239}$Pu is expressed (n,2n$_k$) 
\begin{eqnarray}
{\sigma^{J^\pi}_{n,n',n''_k}}(E_n)=  \sigma_{n}^{CN}(E_n,J^{\pi})   {\frac{\sum _{x\in \Delta E'}  T_{n'_x n''_k}^{J^{\pi}} (E_{n})}  {{T_{tot}^{J^{\pi}} (E_{n})}}}\mbox{ , }
\label{eq:hf2nk}
\end{eqnarray}
with $\Delta E'$ holding the continuum energy range, split in 100~keV bins, from B$_n^{\mbox{\tiny FSRN}}$ to U'*, the maximum excitation energy available in the first-stage-residual nucleus (FSRN). The total (n,2n)  cross section is then,  
\begin{eqnarray}
\sigma_{n,2n}(E_n)= \sum _{J^{\pi}} \sum _{k \in \Delta E''}  {\sigma^{J^\pi}_{n,n',n''_k}}(E_n)  \mbox{ , }
\label{eq:hf2ntot}
\end{eqnarray}
with $\Delta E''$ holding the energy range of the second-stage-residual nucleus from its ground state to the highest energy excited state (or energy bin) possibly reached. Simple formulation of Eq.~(\ref{eq:hf2nk}) may suggest that there is no special comment to be made. However, developing   $T_{n'_x n''_k}^{J^{\pi}}$ using $\mathcal{P}^{(x)}_{n''_k}$ the probability of emitting a second neutron from given first-stage-residual nucleus excitation  energy bin $x$ to the final state of excitation $k$ in the second-stage-residual nucleus, leads to 
\begin{eqnarray}
&&T_{n'_x n''_k}^{J^{\pi}}(E_n)= T_{n'_x}^{J^{\pi}}(E_n) \mathcal{P}^{(x)}_{n''_k} \mbox{ }\\
&&\mbox{ with} \sum _{c''\in n,f,\gamma} \mathcal{P}_{c''}^{J'^{\pi'}}(E_{c''})=\mathds{1} \mbox{  }\mbox{  }\mbox{  }\forall J'^{\pi'} \in \mbox{FSRN} \nonumber\mbox{, }
\label{eq:tnnk}
\end{eqnarray} %$B_n^{^{239}Pu}
expression of which the neutron transmission coefficients are calculated on the model of the Eq.~(\ref{eq:tn}). We emphasize that an energy bin is defined  by its lower and upper energy boundaries, spin and parity values. Choice has been made to compute $\mathcal{P}$  by Monte Carlo (with the {\footnotesize $MC$} label) because it overlaps the level fluctuating energy range of the $^{239}$Pu compound system; meaning the range from 0 to 1.6~MeV in neutron energy scale or equivalently 5.65 to 7.25~MeV  in excitation energy scale (Fig.~\ref{fig:Pgfi239*}). Special notation {\footnotesize $MC-s$}, used in the next equations, according to the calculated Monte Carlo probabilities refers to a simplification~\cite{bou:19} made in present work for the calculation of the width fluctuation correction factors since very little correlation amount is expected between the first step entrance channel (the incident neutron) and  second step outgoing channel (the second-chance neutron). Only exit channel width fluctuation correlations across flux conservation are corrected; meaning the correlations among second-chance reaction decay widths distributions.

\begin{figure}[t]
\center{\vspace{1.0cm}
\resizebox{0.99\columnwidth}{!}{
\includegraphics[height=5cm,angle=0]{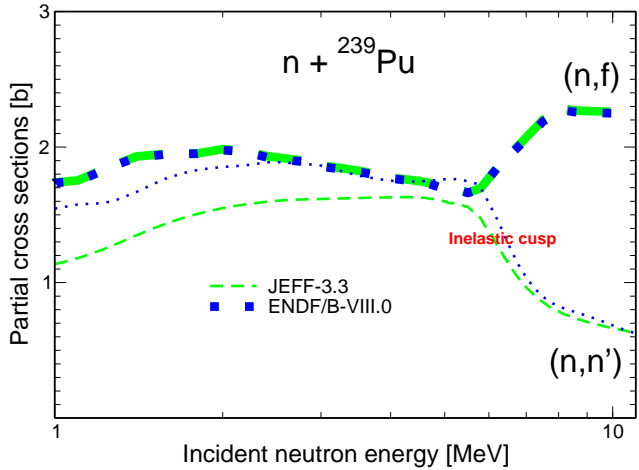}} }
\caption{\label{fig:cusp}(Color online) $^{239}$Pu neutron-induced total inelastic and total fission cross sections evaluated in the framework of ENDF/B-VIII.0~\cite{bro:18} (blue short dashed lines) and JEFF-3.3~\cite{plo:20}) (green long dashed lines)  projects. Thick and thin curves correspond respectively to the $(n,f_{tot})$ and $(n,n'_{tot})$ reactions.}
\label{fig:239cusp}
\end{figure}

\subsubsection{Depleted first-chance inelastic cross section}
The sudden opening of second-chance reactions about the neutron binding energy (B$_n^{239}$=5.65~MeV) contributes to the Wigner-cusp-type~\cite{wig:48} inflection observed in the $^{239}$Pu neutron-induced total inelastic scattering cross section as exemplified by Fig.~\ref{fig:239cusp}. This drop is maximized by a rapid disappearance of compound nucleus mechanisms as energy increases. In our computer code, the flux removal from the CN inelastic cross section due to the  second-chance reactions opening is applied on the 'first-chance-only' inelastic transmission coefficient, $T_{n'_x}^{J^{\pi}}$. The expression of the depleted  transmission coefficient for given spin-parity is therefore,  
\begin{eqnarray}
T_{n'_x}^{depl.,J^{\pi}}(E_n)&=& T_{n'_x}^{J^{\pi}}(E_n) \nonumber\\ 
&-& T_{n'_x}^{J^{\pi}}(E_n) \sum _{k \in \Delta E''} \mathcal{P}^{x}_{n''_k;MC-s}  \nonumber\\
&-& T_{n'_x}^{J^{\pi}}(E_n)  \mathcal{P}^{x}_{f'';MC-s} \nonumber\\
&-& T_{n'_x}^{J^{\pi}}(E_n)  \mathcal{P}^{x}_{\gamma'';MC-s} \mbox{ .}
\label{eq:tdep1}
\end{eqnarray}
Replacing  $T_{n'_x}^{J^{\pi}}$ by $T_{n'_x}^{depl.,J^{\pi}}$  in the pure Hauser-Feshbach equation, returns the correct value of the CN inelastic  partial cross section, $\sigma^{CN, depl.,J^\pi}_{n,n'}$. Remembering Eq.~\ref{eq:totinel1}, the expression of the total inelastic cross section is reduced to,
\begin{eqnarray}
\sigma_{n,n'}=  \sigma_{n,n'}^{direct-disc.} + \sigma_{n,n'}^{PE} + \sum _{J^{\pi}}\sigma^{CN, depl.,J^\pi}_{n,n'}\mbox{ ,  }
\label{eq:totinel2}
\end{eqnarray}
where the incident neutron energy dependence, $E_n$, has been left aside for display.

\subsubsection{\label{sss:estimate}Second-chance preequilibrium neutron emission}

The moment has come to remember the existence of two distinct neutron emission paths leading to the 'same' residual nucleus meaning by direct and compound nucleus interactions. Direct interactions with residual nucleus left in a low-lying  level can not trigger a second-chance neutron emission whereas direct interactions with residual nucleus left in a state of the neutron continuum above the neutron emission threshold can contribute to the (n,2n) cross section (Fig.~\ref{potWells}). Preequilibrium (PE) theory supplies the amount of inelastic scattering due to the latter contribution in the total inelastic cross section. PE inelastic reactions lead to a non-equilibrated residual nucleus, meaning in a state of excitation high in the continuum but formed by a very few number of particle-hole excitations (indiscriminately referred to as {\it excitons} in the exciton model). Possibly, one of the excited neutron-particle state may decay by neutron emission. It is conventionally described as {\it second-chance preequilibrium neutron emission}. In present work, the fraction of the (n,2n) cross section caused by second-chance preequilibrium has been estimated on the following basis. Starting by a TALYS pre-calculation of $\sigma_{nn'}^{PE}  (E_n)$, the  preequilibrium total inelastic cross section as a function of neutron energy  according to the (n+$^{239}$Pu) reaction, we share  $\sigma_{nn'}^{PE}  (E_n)$ among all $x$ possible  residual nucleus  discrete states (or bins) according to $\rho'_{N}$, the corresponding normalized level density. By assuming second-chance neutron emission probabilities $\mathcal{P}_{n''}$ analogous to CN second-chance neutron emission probabilities, it reads for a  preequilibrium partial cross section ranging from $E_{min}$ (TALYS default is set to $1$~MeV) to $E_n$, the neutron incident energy,    
\begin{eqnarray}
 \sigma_{n,n'_x}^{PE}(E_n)  = \sigma_{n,n'}^{PE}  (E_n)  \rho'_N(x) \mbox{ with,  }
\label{eq:diffPE1}
\end{eqnarray}
\begin{eqnarray}
\rho'_N(x)  = \frac{\rho'(x) }{ \sum_{x \in ([E_n-E_{min}], \mbox{\tiny any } J^{\pi}) }\rho'(x)}  \mbox{  }  \mbox{   }\mbox{ .  } 
 \label{eq:diffPE3}
\end{eqnarray}
The above is made on the assumption that 
\begin{eqnarray}
 \sigma_{n,n'}^{PE}  (E_n) &\equiv&  \sum_{ J^{\pi} } \int_{E_{min}}^{E_n} dE'\mbox{ } \sigma_{n,n'_{x}}^{PE} (E_n) \\
&\approx&   \sum_{x \in ([E_n-E_{min}], \mbox{\tiny any } J^{\pi}) }\sigma_{n,n'_{x}}^{PE} (E_n) \mbox{.} \nonumber
\label{eq:diffPE2}
\end{eqnarray}
Finally, the total  (n,2n)  cross section preequilibrium component (using second-chance neutron probabilities of Eq.\ref{eq:tnnk}) is recovered as,
\begin{eqnarray}
\sigma_{n,2n}^{PE}(E_n)= \sum_{k \in \Delta E''}\sum_{x \in \Delta E'}  \sigma_{n,n'_x}^{PE}(E_n)\mathcal{P}^{x}_{n''_k;MC-s} \mbox{ .} 
\label{eq:diffPE4}
\end{eqnarray} %$B_n^{^{239}Pu}
Once more the energy bin reached in the first-stage-residual nucleus is restrained to $\Delta E'$ (as defined by Eq.~\ref{eq:hf2nk}). The above  total (n,2n)  cross section  preequilibrium component, added to the previously calculated total (n,2n)  cross section compound nucleus component, is consistently subtracted from the total inelastic cross section of Eq.~\ref{eq:totinel2}. 

\subsection{\label{combi}Macro-microscopic level density calculations}
Equation~\ref{eq:diffPE1} opens the question of the most relevant  type of level density ($\rho'$) for  second-chance preequilibrium neutron emission calculation. According to TALYS, the spin-dependent population calculated after compound nucleus emission ({CNLD}) works well with the exciton model~\cite{gad:92} whereas the MSD-MSC quantum mechanical preequilibrium model is more adequately represented  by a spin distribution based on particle-hole state densities. Several types of residual nucleus level density have been tested in this work using the AVXSF-LNG code features, including the total level density. Results carried by the CNLD will be compared to those involving non-equilibrated residual nucleus level densities later in this paper. The CNLD in our computer program is simply reconstructed from HF neutron emission  transmission coefficients, with similar normalization to Eq.~\ref{eq:diffPE3} except that it can now be discriminated per $J^\pi$. An additional normalization coefficient, $N_{Jls}= 1/\big[\sum_{J^\pi(l,s)}\mathds{1}\big]$, is then necessary to preserve the preequilibrium flux supplied by TALYS according to the sum of $J^\pi$ states populated; meaning    
\begin{eqnarray}
\rho'_N(x)=   N_{Jls} \frac{T_{n'_x}^{J^{\pi} }(E_{n})}  { \sum_{x \in ( [E_n-E_{min}], J^{\pi}) }{T_{n'_x}^{J^{\pi} }(E_{n}) } } \mbox{  .}
\label{eq:cnld}
\end{eqnarray}
As far as a spectrum simulation of a few number of individual excitations is wanted,  the combinatorial Quasi-Particle-Vibrational-Rotational (QPVR) Level Density (LD) method (section  III of Ref.~\cite{bou:13}) implemented in the AVXSF-LNG computer program  is well suited to reach that goal.  In contrast to the exciton method available in the TALYS code, our phenomenological model is based on the concept of {\it excited quasi-particle states} for the individual excitations. We start from the formula for  the quasi-particle energy $E_\nu^{exc}$, in terms of the independent particle states of a Nilsson-type spectrum in a deformed potential well with $\Delta$, the pairing energy parameter. The latter being taken not from the solution of the full pairing equations~\cite{row:70}  but as a parameter value evaluated from experiments. It reads for the quasi-particle energy equation, meaning for the excitation energy of a quasi-particle resulting from the breaking of a nucleon pair in the state $\nu$ with $e_\nu$ the nucleon orbital energy,
\begin{eqnarray} 
E_\nu^{exc}=\sqrt{(e_\nu-\lambda)^2+\Delta^2}\label{eq:qpenergies}
\end{eqnarray}
 where $\lambda$ plays the role of the Fermi energy  (with differentiation between neutron and proton, and a present Nilsson states database currently set up from  Ref.~\cite{mol:16}). Present combinatorial strategy is the following: the energies of multi-quasi-particle states are established additively whereas collective states are formed by adding the know or estimated values of vibrational excitations of various kinds ($\beta_2$ prolate, $\gamma$, mass asymmetry octuple and bending vibrations)  and finally,   rotational bands are constructed on top of the previously set up band heads.\\

\begin{figure}[t]
\center{\vspace{0.4cm}
\resizebox{0.99\columnwidth}{!}{
\includegraphics[height=5cm,angle=0]{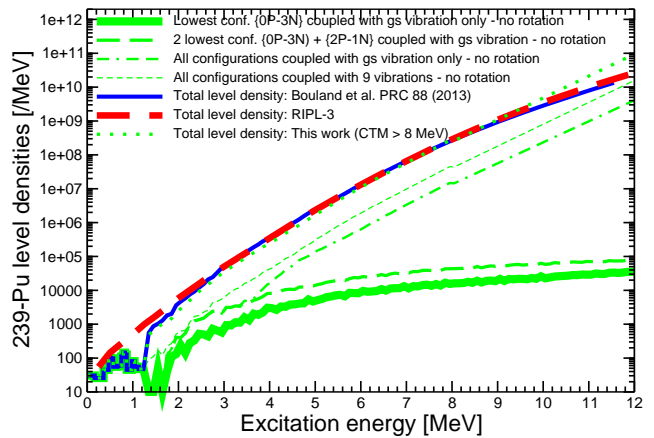}} }
\caption{\label{fig:QPVRLD239}(Color online) Components of the total level density  as a function of excitation energy at ground state deformation in the $^{239}$Pu* residual nucleus. Displayed simulations correspond to, from the bottom  to the top,  this work, the microscopic HFB-BCS~\cite{cap:09,gor:09} and the 2013 database~\cite{bou:13} total level densities. Then follows some  QPVR simulations successively with no rotational enhancement, no  vibrational enhancement, no multi-quasi-particle states except the \{0P-3N\} and \{2P-1N\} lowest excited configurations and finally with simply  the \{0P-3N\} lowest excited configuration.}
\end{figure}

We recall that in quasi-particle theory (the Bardeen-Cooper-Schrieffer (BCS) theory~\cite{bar:57}) there are no {\it holes} and {\it particles}, only {\it quasi-particles}. In the ground state the pairing forces combine all the particles into a kind of {\it vacuum} state, lowering the energy below that of the equivalent state in the independent particle model. In this {\it pair-correlated ground state}, the orbitals near the Fermi energy have about 50 \% pair occupancy and as one goes down in energy this occupancy probability increases until it is near 100\% at about $\lambda - \Delta$. Conversely, the pairing occupancy probability decreases as one goes above the Fermi energy, becoming close to zero at $\lambda + \Delta$. For an even-even (e-e) nucleus, there are no excited quasi-particles in the ground state. The lowest excited quasi-particle states (of two-quasi-particles type) are formed by breaking a pair of neutrons or protons giving a total excitation energy of 
\begin{eqnarray} 
E^{exc}_{2qp}=E^{exc}_\nu+E^{exc}_{\nu'}\label{eq:2qp}
\end{eqnarray}
on the model of Eq.~\ref{eq:qpenergies}. When $(e_\nu-\lambda)$ and $(e'_\nu-\lambda)$ are small compared with $\Delta$ there is an energy gap of just over $2\Delta$ between the ground state and the lowest quasi-particle states.  These two-quasi-particle states can be pictured as the removal of two orbitals from the pair-correlated system with additional energies calculated as unpaired particles in these orbitals. The excited multi-quasi-particle  states in the program are identified in terms of configurations such that two-neutron quasi-particle states are  labelled as \{0P-2N\}. Regarding the odd-mass $^{239}$Pu residual nucleus, the ground state comprises a one-neutron quasi-particle state excitation, equivalently described as \{0P-1N\}. Since the corresponding neutron pairing energy $\Delta_n=0.6979~MeV$, is lower~\cite{mol:16} than the proton pairing energy, $\Delta_p=0.8577~MeV$, the lowest excited multi-quasi-particle  states are expected to belong to the \{0P-3N\} configuration. On the same footing,  the next excited multi-quasi-particle  states sequence corresponds to the \{2P-1N\} configuration. The cumulative sum of the expected configurations to the total level density  for the $^{239}$Pu residual nucleus is drawn in Fig.~\ref{fig:QPVRLD239}, starting by the lowest excited configurations  \{0P-3N\} and  \{2P-1N\}  and concluded by  the sum of all sets of quasi-particle states  excitations. Later  individual excitations full sequence is  combined with multi-phonon states excitations (listed in Table II of Ref.~\cite{bou:18}) to form band heads. The whole spectrum of model states is finally completed by building  the rotational band associated to each band head according to the quantum numbers involved (Eq.~(44) of Ref.\cite{bou:13}).\\ 

Consequences on the $^{239}$Pu(n,2n) cross section of the choice made for the residual nucleus level density, as simulated with the QPVR module or using  Quasi-particle Random Phase Approximation (QRPA) results~\cite{dup:17},  is discussed at the end of this paper. 

%Regarding valuable estimation of the second-chance preequiibrium neutron emission (paragraph~\ref{sss:estimate}) using Eq.\ref{eq:diffPE4}, impact of the \{0P-3N\} lowest excited states level density (thickest-solid-green curve on Fig.~\ref{fig:QPVRLD239}) has been evaluated against the spin-dependent population after compound nucleus emission (Eq.~\ref{eq:cnld}).    

%-- FIGURE -- LD Vs RIPL 

\section{\label{s:application}Evaluation of the (\MakeLowercase{n}+$^{239}$P\MakeLowercase{u}) reaction cross sections over the [1~keV-12~MeV] energy range} 
\subsection{\label{ss:total} Total CN formation cross section assessment}
Following our strategy to carry a unique and consistent nuclear database according to the Pu isotopes, we start this work by total cross section fitting according to the (n+$^{238}$Pu) and (n+$^{239}$Pu) reactions respectively up to second- and third-chance reaction thresholds. This is  achieved across an ECIS-06 CCOM calculation, driven by  the TALYS-1.95 software, that  couples for both target isotopes the ground state and the fifth lowest rotational levels using the global optical model parameters (OMP) of Soukhovitskii \etal~\cite{sou:04}  dedicated to actinides together with the nucleus masses and, $\beta2$ and $\beta4$  deformation parameters tabulated by M\"oller \etal~\cite{mol:95}. It turned out that the global OMP sets work satisfactorily for both target isotopes.\\   

\begin{figure}[]
\center{\vspace{1.0cm}
\resizebox{0.99\columnwidth}{!}{
\includegraphics[height=5cm,angle=0]{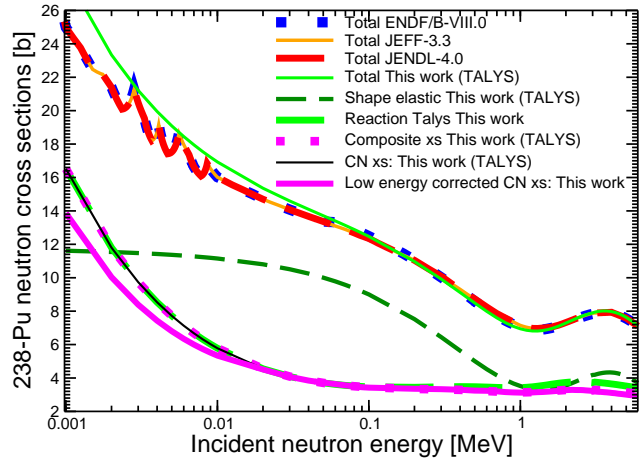}} }
\caption{\label{fig:pu238tot}(Color online) Comparison of the neutron-induced total cross section computed with the TALYS software and compared to some evaluated results by ENDF/B-VIII.0~\cite{bro:18}, JEFF-3.3~\cite{plo:20} and   JENDL4.0~\cite{jendl4u} according to the $^{238}$Pu target nucleus. The subdivision of the total cross section into the shape elastic and reaction ($\sigma^{R}$) cross sections is also addressed. Subtracting the shape inelastic (i.e.; the direct inelastic) and the preequilibrium cross sections from $\sigma^{R}$ supplies the total compound nucleus formation cross section needed by the AVXSF-LNG code to perform HF calculations. }
\end{figure}

\begin{figure}[]
\center{\vspace{1.0cm}
\resizebox{0.99\columnwidth}{!}{
\includegraphics[height=5cm,angle=0]{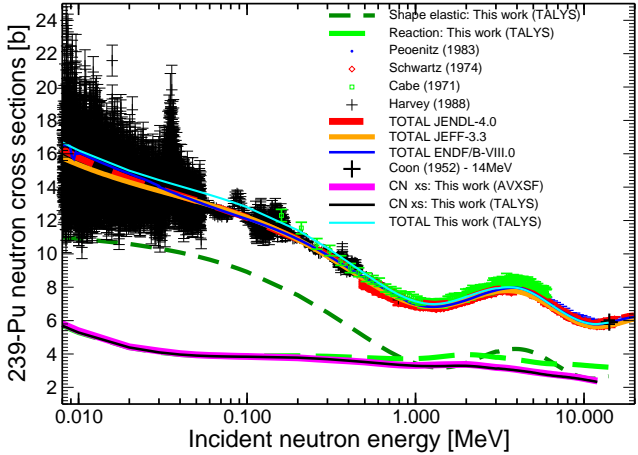}} }
\caption{\label{fig:pu239tot}(Color online) Similar to Fig.~\ref{fig:pu238tot} but according to the $^{239}$Pu target nucleus.}
\end{figure}

Since there is no total cross section measurements above 200~eV for the (n+$^{238}$Pu) reaction, benchmarking of present work involves a comparison with recommended evaluations. On the opposite the $^{239}$Pu target isotope database includes a large choice of total cross section  measurements (derived from transmission measurements) in the neutron continuum by Schwartz \etal~\cite{sch:74}, Coon \etal~\cite{coo:52}, Poenitz \etal~\cite{poe:83}, Harvey \etal~\cite{har:03} and Cabe \etal~\cite{cab:71}. Figures~\ref{fig:pu238tot} and~\ref{fig:pu239tot} show the good quality of the fits respectively for the two target isotopes with corresponding measurements and evaluated curves. According to our  (n+$^{239}$Pu) total cross section calculation (Fig.~\ref{fig:pu239tot}), there are two differences with the JEFF-3.3 recommendation.  First, below 600~keV present TALYS calculation is likely too high  because it follows the old data by Cabe \etal~\cite{cab:71} and not the more recent and accurate transmission measurement by Harvey \etal~\cite{har:03}. On the other side, JEFF-3.3 is lower at 14~MeV than present calculation that follows consistently with ENDF/B-VIII.0~\cite{bro:18} and  JENDL4.0u~\cite{jendl4u} the experimental data point published by Coon \etal~\cite{coo:52}. We then expect JEFF-3.3 at high energy to predict a lower direct inelastic cross section than the others.\\

\begin{figure}[]
\center{\vspace{1.0cm}
\resizebox{0.99\columnwidth}{!}{
\includegraphics[height=5cm,angle=0]{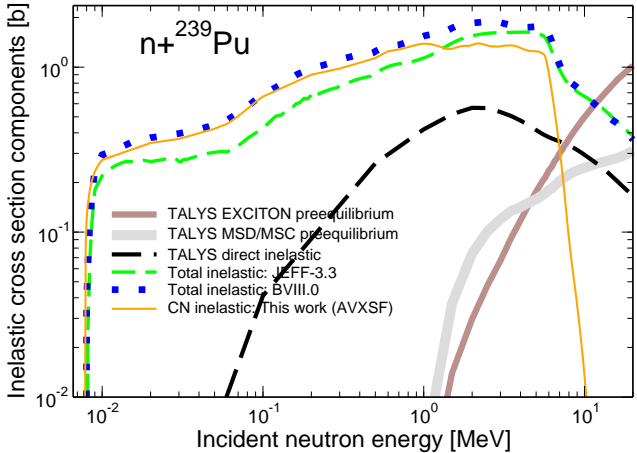}} }
\caption{\label{fig:preequi}(Color online) Preequilibrium model comparison for the (n+$^{239}$Pu) reaction below 20~MeV. Comparison of the MDS/MSC and  exciton  options as proposed by TALYS. The total inelastic cross sections recommended by  ENDF/B-VIII.0~\cite{bro:18} and JEFF-3.3~\cite{plo:20} deviate from each other below 6~MeV.}%Present total inelastic cross section calculated is compared to the ENDF/B-VIII.0~\cite{bro:18} and JEFF-3.3~\cite{plo:20} recommendations.}
\end{figure}

At neutron energies below first inelastic threshold, respectively at 7.8 and 44~keV for the $^{239}$Pu and $^{238}$Pu residual nuclei, there is a strict equivalence between the total compound nucleus cross section  and the reaction cross section ( $\sigma^{CN}\equiv\sigma^{R}$ ). Within the energy region between the inelastic threshold and the energy from which  preequilibrium  becomes significant (around 1~MeV), $\sigma^{CN}$  is equivalent to the composite-nucleus formation cross section  (Eq.~(\ref{eq:totcn})). Above the preequilibrium threshold energy, the extraction of $\sigma^{CN}$ requires the  derivation of the full set of equations defined in  paragraph~\ref{s:over}. As mentioned before, two models are available to treat preequilibrium in   TALYS released  versions: a  two-component exciton classical model~\cite{kon:12} and a multistep direct-multistep compound (MSD/MSC) quantum mechanical model~\cite{kon:97}. Figure~\ref{fig:preequi} shows for the (n+$^{239}$Pu) reaction below 20~MeV, a comparison between the two   preequilibrium modeled cross sections by TALYS in terms of magnitude and shape relatively to the amount of total inelastic scattering (Eq.~(\ref{eq:totinel1})); the latter commonly calculated as the sum of the CN, direct inelastic and preequilibrium components. Although  preequilibrium profiles remain close for the two models over the threshold energy range, the two calculations return different shapes above 4.5~MeV with a much stronger 'exciton-based' angular-integrated cross section magnitude (3 times larger at 20~MeV). We guess that the choice of one model over another will conduct to a significant change in the extracted $\sigma^{CN}$ profile at  high energies. Since above 1~MeV neutron energy, CN radiative decay and CN elastic scattering are negligible, any change in $\sigma^{CN}$ will be absorbed essentially by the  fission parameters  since neutron strength functions are constrained by the entrance flux (meaning the total compound nucleus formation cross section). As far as the (n+$^{238}$Pu) reaction is concerned, the value of the preequilibrium angular-integrated cross section reaches about 150~mb at $B_n^{^{239}Pu}$=5.65~MeV irrespective of the TALYS model used. We may state that present (n+$^{238}$Pu) results, performed at the energies lower than $B_n^{^{239}Pu}$, are independent of the preequilibrium model  selected.\\ 
     
Regarding possible weakness of present (n+$^{238}$Pu) low energy total cross section fit ($E_n<100$~keV on Fig.~\ref{fig:pu238tot}) that could be propagated into the calculation of the  $^{239}$Pu decay probabilities,  a correction is applied on the l-wave strength functions ($S_l$; Appendix~\ref{connect}), to return the most valuable estimation of the total compound nucleus cross section. The $\sigma^{CN}$ correction at low energy relies on best estimates of the $S_0$ and $S_1$ asymptotic values, of about $1.044\times10^{-4}$ and $1.48\times10^{-4}$, as suggested by our previous  study~\cite{bou:13} encompassing the whole Pu isotope series database. Differences between the TALYS predicted curve and the  revised shape according to $\sigma^{CN}$  can be checked on Fig.~\ref{fig:pu238tot} (purple thick-dotted line vs. purple thick solid curve). For the (n+$^{239}$Pu) reaction, the total cross section profile around 100~keV shows possible improvement but no significant correction was applied to  $\sigma^{CN}$ since in present article we are mainly concerned by second-chance reactions. However it must be noted that our calculated total cross section  converges  to the ENDF/B-VIII.0~\cite{bro:18} and  JENDL-4.0~\cite{jendl4u} evaluations as neutron energy decreases (Fig.~\ref{fig:pu239tot}).

\subsection{\label{ss:fissionLD}Macro-microscopic vs. HFB-BCS level densities  }

\begin{figure}[]
\center{\vspace{1.0cm}
\resizebox{0.99\columnwidth}{!}{
\includegraphics[height=5cm,angle=0]{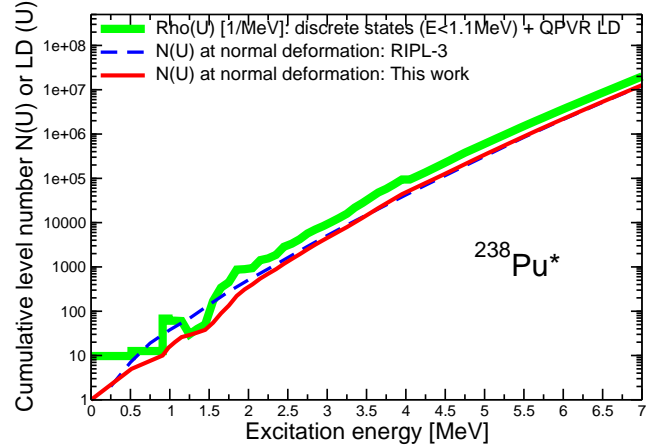}} }
\caption{\label{fig:238ldnd}(Color online) $^{238}$Pu* cumulative number of levels as a function of excitation energy at ground state deformation constructed in this work and compared with the HFB-BCS prediction~\cite{gor:09}. Present work QPVR level density [1/MeV] including low-lying discrete levels observed~\cite{dim:20} or expected up to 1.1~MeV, is drawn for later discussion.}
\end{figure}

\begin{figure}[]
\center{\vspace{1.0cm}
\resizebox{0.99\columnwidth}{!}{
\includegraphics[height=5cm,angle=0]{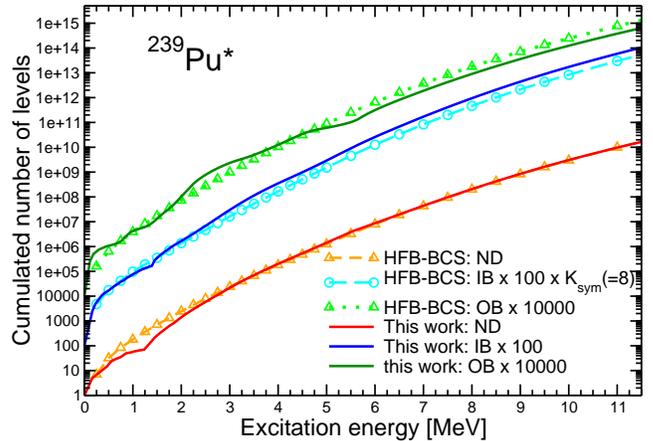}} }
\caption{\label{fig:239ld}(Color online) Present work $^{239}$Pu* cumulative number of levels as a function of excitation energy at ground state (ND), first saddle (IB) and second saddle (OB) deformations compared with corresponding HFB-BCS predictions~\cite{gor:09}. Curves displayed correspond, from the bottom to the top, to ground state, inner (multiplied by 100) and outer  (multiplied by 10000) barrier deformations. To ease present results comparison at the inner barrier, a maximum enhancement factor of  $8$ for axial symmetry breaking (triaxiality) with mirror symmetry, was applied to RIPL-3.}
\end{figure}

\begin{figure}[]
\center{\vspace{1.0cm}
\resizebox{0.99\columnwidth}{!}{
\includegraphics[height=5cm,angle=0]{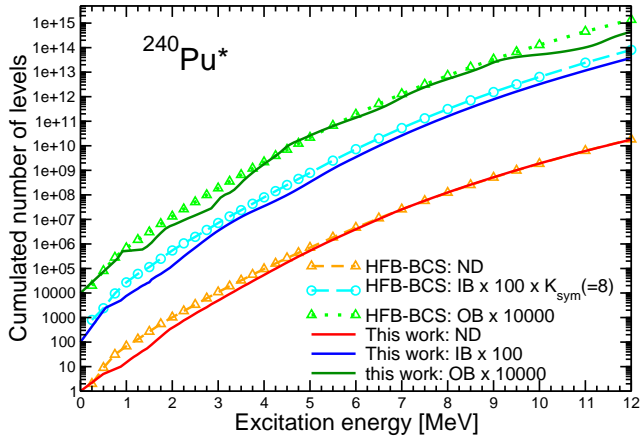}} }
\caption{\label{fig:240ld}(Color online) Same as Fig.~\ref{fig:239ld} but for $^{240}$Pu*.}
\end{figure}

In our 2013~\cite{bou:13} analysis carried over the whole Pu isotope series, special attention was paid to modeling level density functions over the neutron energy range up to second-chance fission threshold. The level densities (LD) fitted previously according to normal deformation (ND), inner and outer fission barrier deformations were used as prior knowledge in this work but also extended to upper energies according to the (n+$^{239}$Pu) reaction. Present LD assignments are also supported by our macro-microscopic combinatorial level density model (QPVR LD model), especially at normal deformation, for which nucleon pairing energy parameters were adjusted to reproduce the observed slow neutron average resonance spacing values as listed in Table III of Ref.~\cite{bou:13}. Sequences of low-lying states are set up for the target nuclei from the Evaluated Nuclear Structure Data File (ENSDF)~\cite{dim:20}, augmented where it seems likely by additional levels to complete the rotational bands as expected by our model. The consistency between input parameter and nuclear structure data was ensured among all Pu compound nuclear systems involved, for which level structures are especially dependent on the e-e or even-odd (e-o)  character (Fig.~2 of Ref.~\cite{bou:20}). The former category implies only collective states at low energy (Eq.~\ref{eq:2qp}) on the opposite to e-o (or o-e) systems which carry an inset of low energy quasi-nucleon states. At respectively 1.1, 1.3 and 1.3~MeV excitation energies above  ground state for the $^{238}$Pu, $^{239}$Pu  and $^{240}$Pu target nuclei, precise knowledge on  individual states fails. Our in-house QPVR method supplies then a confident alternative to build the requested level densities. Our model relies, in particular, on  best evaluation of the neutron ($n$) and proton ($p$) pairing energies ($\Delta$ in Eq.~(\ref{eq:qpenergies})) as well as $\Im$, the moment of inertia; all those parameters being dependent on nucleus deformation and excitation energy. Strong considerations on those parameters are difficult to extract from neutron-induced cross-section experimental data fits alone, but the present work is aiming to marry theoretical and experimental information. Figure~\ref{fig:238ldnd} displays the cumulative number of levels, built using our QPVR model, corresponding to the ground state of the $^{238}$Pu with pairing energy values $\Delta_n$=0.70~MeV and $\Delta_p$=0.78~MeV and with $\Big[\hbar^2/(2\Im)\Big]$=6.5~keV. 
%valeurs dans doc. Eric
These values are from a global expression for pairing energy parameters drawn from second order mass differences ($\Delta=12/\sqrt A)$, with a 10\% reduction for the lower trend of $\Delta_n$, in the actinides region. Level densities  built and adjusted in this work are  systematically compared to the Hartree-Fock-Bogolyubov - Bardeen-Cooper-Schrieffer (HFB-BCS) predictions as addressed by Goriely  \etal~\cite{gor:09} (as exemplified by Figs.~\ref{fig:238ldnd}, ~\ref{fig:239ld} and~\ref{fig:240ld}).\\

% Our fitted multiphase LD, $ \rho(U,T)$, by using a constant but independent temperature $T$ in each phase, 
At low excitation energies above fundamental fission barriers, ad hoc sequences of individual transition states were generated using again the  QPVR model, since  direct observation of those levels is quite difficult.  As incident neutron energy increases, detailed resonance structure is of much less importance and the fission cross-section  depends mainly on the level densities of the compound nucleus at barrier deformations. After some trial-and-error, we  decided to use our QPVR model for the inner barrier whereas an empirical approach for the outer barrier in the form of the multiphase-temperature level density model (Eq.~(10) of Ref.~\cite{bou:13}) was chosen. Latter model parameters were tuned to reproduce at best experimental fission cross section data, but in  preserving the trend carried by the prior QPVR calculation. In that sense the multiphase-temperature model (MCTM) parameters tuned in this work, are our main fitting ingredients that absorb all deficiencies carried by present modeling; meaning in particular the unique one-dimensional fission path and the representation of fission barriers as inverted parabola and finally  possible non-statistical effects observed in the measured cross sections. We emphasize that use of temperature-dependent LD has been practiced also for adjusted  microscopic HFB calculations~\cite{hil:10} but over wider energy range (0-200 MeV). Figures~\ref{fig:239ld} and~\ref{fig:240ld}  show the cumulative number of levels at ground state and, at inner and outer barrier deformations that are obtained in the present study respectively for the $^{239}$Pu*  and $^{240}$Pu* excited nuclei, including the HFB prediction of Ref.~\cite{gor:09}. The RIPL-3 LD at second saddle contains by default the left-right asymmetry enhancement factor ($K_{sym}\approx 2$) whereas the corresponding first saddle data do not contain any asymmetry enhancement factor. Thus for actual comparison in terms of inner barrier cumulative number of levels, a maximum enhancement factor of  $8$ for axial symmetry breaking (triaxiality) with mirror symmetry, was applied to RIPL-3 data at first saddle. A comparable value was  extracted from systematics on the Np family by G. Vladuca \etal~\cite{vla:06}.\\

Generally speaking, the shape and magnitude of cumulative number of levels at ground state and at inner barrier deformation  are in quite reasonable agreement. In terms of outer barrier, magnitudes are similar although the wavy shape of our temperature-dependent LD model~\cite{bou:13}  departs from the monotonic slope of the RIPL-3 prediction. Since barrier heights and transition state level densities play an anti-correlated role, there are several possible combinations of input parameters that results to similar degrees of confidence (here no better than $\pm$200 keV in terms of barrier heights). By comparison to the pure microscopic HFB barrier values listed in Table~\ref{tab:barriers} that are higher by several hundreds of keV according to the $^{240}$Pu fissioning nucleus, we expect consistently higher level densities. On the other side, the authors of study~\cite{hil:10} have logically obtained a significant agreement improvement by normalizing individually prior HFB fission barriers (prior calculated fission cross sections were too low in magnitude) or an even better compatibility by fine-tuning the LDs, similarly to present approach, on both low-lying levels and observed neutron resonance spacings at $B_n$.

%Present $\sigma_{n,n'}$ according to the  (n+$^{239}$Pu)  system lies in between the ENDF/B-VIII.0 and the JEFF-3.3 curves whereas the same reaction for the (n+$^{238}$Pu) system deviates from the two evaluations above 200~keV. In our 2013~\cite{bou:13} analysis, a special attention was paid to modeling level density functions and in particular to the level density of the target nucleus at normal deformation, which controls the competing inelastic neutron scattering reaction. The latter level density, built using our semi-microscopic combinatorial level density model~\cite{bou:13}, is part of the differences observed. 

\subsection{\label{s:partials}Compound nucleus partial cross sections}

\subsubsection{\label{ss:general}General considerations}

In section~\ref{s:app}, we have presented our methodology, starting by CCOM calculations to reproduce (or predict when there is no experimental data) total cross sections according to the (n+$^{238}$Pu) and (n+$^{239}$Pu)  systems. Fits on the former system supply the set of input parameters needed for accurate computation of the decay probabilities of $^{239}$Pu* (introduced by Fig.~\ref{fig:Pgfi239*}) and subsequently related second-chance cross sections. Among sensitive items in the calculation of the (n,2n) cross section  are, by order of appearance, 
\begin{enumerate}
\item the choice of  preequilibrium model,
\item the selection of the best experimental fission cross section data set that would make reference at above 5~MeV neutron energy ,
\item the most physical (inelastic) level density according to the second-stage-residual nucleus at below 2.5~MeV excitation energy and, 
\item the type of channel width fluctuation correction factor to be used in the HF decay probability equations at below 1.6~MeV excitation energy.  
\end{enumerate}
Regarding items 1 and 2), among the two TALYS alternatives tested for flux leakage due to preequilibrium reactions prior to CN decay in the (n+$^{239}$Pu)  reaction, the MSD/MSC results have led in present work to an unreasonable outer barrier  MCTM profile at high neutron energy according to the  $^{240}$Pu fissioning nucleus. However if we put this issue aside, it is actually the shape and magnitude of the measured $^{239}$Pu neutron fission cross section making reference which is important because it governs by reciprocity the magnitude of the first-stage-only  total inelastic scattering HF transmission coefficient, $T_{n'_x}^{J^{\pi}}$ of Eq.~(\ref{eq:tdep1}) (second-chance fission parameters unvaried), and finally the HF contribution to the (n,2n) cross section. To conclude this paragraph and on the fact that an unreasonable outer barrier profile at high neutron energy was obtained in present work when using the MSD/MSC model, {\em the results selected hereafter are based on the exciton preequilibrium model}. However, for global uncertainties feedback, an alternative calculation using the MSD/MSC component is presented in Appendix~\ref{unc}. \\

Item 3) We expect the level density in the $^{238}$Pu second-stage-residual nucleus to play a significant role in present case  since  latter nucleus carries an even number of neutrons and protons. The low-lying energy spectrum of e-e nuclei is built solely from pure collective  excitations (as illustrated by Fig.~2 of Ref.~\cite{bou:20}) by contrast to odd and o-o nuclei until a first pair of nucleons  be broken at an excitation energy equal to two times the nucleon pairing energy (as suggested by Eq.~\ref{eq:qpenergies}). Assuming neutron and proton pairing energy values of respectively 0.70~MeV and 0.78~MeV, discontinuities would show up in the $^{238}$Pu* level density spectrum when the excitation energy crosses $2\Delta_n$, $2\Delta_p$, $2\times2\Delta_n$, $2\times2\Delta_p$ meaning at about 1.40, 1.56, 2.80 and 3.12~MeV. However 2 quasi-particle states may be observed at an excitation energy lower than the one expected   (at $2\Delta_n$ or $2\Delta_p$) as it seems to be in the $^{238}$Pu* nucleus system with the report in the ENSDF database~\cite{dim:20} of a $4^-$  2-quasi-particle state at about 1.083~MeV (see also Tab.~\ref{tab:lowLying} as input in AVXSF-LNG). The peculiar low energy spectrum pattern of the $^{238}$Pu e-e nucleus, illustrated by  Fig.~\ref{fig:238ldnd}, deviates from the smooth behaviour carried by the RIPL-3 LD curve up to  3~MeV excitation energy. This 2-quasi-particle state specificity has been also pointed out by Maslov~\cite{mas:21} when studying prompt fission neutron spectra. \\ 

\begin{figure}[]
\center{\vspace{1.0cm}
\resizebox{0.99\columnwidth}{!}{
\includegraphics[height=5cm,angle=0]{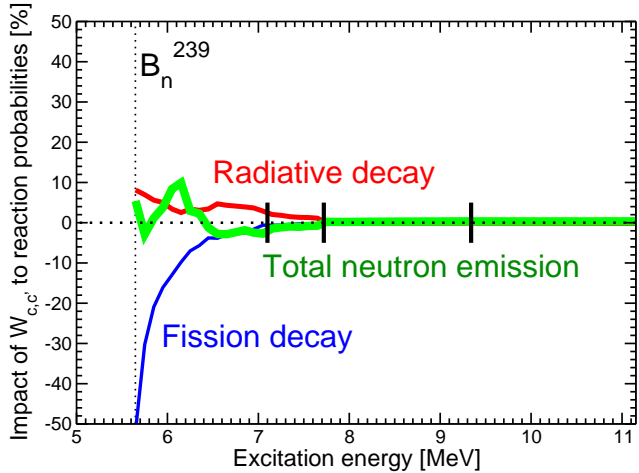}} }
\caption{\label{fig:wnow}(Color online) Error [\%] expected when no surrogate-dedicated channel width fluctuation correction factor  (SWFCF) is made when calculating the $^{239}$Pu* decay probabilities. Black vertical lines correspond respectively to the  energies of the (n,2n)  cross section measurement published by M\'eot \etal~\cite{meo:21} at respectively 7.1, 7.72 and 9.34~MeV neutron energy.}
\end{figure}

Concerning item 4), we mentioned as preamble that present work does not carry the Engelbrecht-Weidenm\"uller (EW) transformation but it was reported by Kawano \etal~\cite{kaw:16} that  latter approximation investigated over the [(0-1)~MeV] neutron fluctuating energy range,  for actinides, can propagate  up to 7\% change according to the inelastic channel since only a few reaction channels are open. Back to present study for the calculation of  reaction decay probabilities, it likely impacts the set of input nuclear parameters according to the fitted (n+$^{238}$Pu) reaction and thus indirectly to the recalculated low-energy HF decay probabilities. In this work, we were only able to quantify the remaining part of the correlation factor, meaning the amount of correlations between the exit reaction channel widths in the CN hypothesis. This topic has been discussed in detail in two recent publications~\cite{bou:19,bou:20} and will not be discussed further in present article. We will just underline that present reaction decay calculation relies on the hypothesis of {\it very little correlation between the first step entrance channel (the in-going neutron impinging the $^{239}$Pu target nucleus) and the second step outgoing neutron (i.e; the 2n exit channel)}. Therefore present HF decay probability calculation is corrected only for expected  correlations between second-chance -fission, -neutron inelastic emission and -radiative decay widths distributions.  Figure~\ref{fig:wnow} returns an estimation of the error brought by the total absence of width fluctuations correction factor (so-called SWFCF in Ref.~\cite{bou:19} ) in the calculation of  compound nucleus decay probabilities. The conclusion of the above SWFCF sensibility study is that channel width fluctuations impact likely only the calculation of the lowest energy data point measured by M\'eot \etal~at 7.1~MeV and this, with reasonable magnitude ($<$2.5\%) compared to other sources of uncertainties. We can make the hypothesis that the EW transformation brings a correction of similar magnitude in the most exact calculation. 

%What differentiates in particular the present procedure from the original TALYS-based approach is the absence of SWFCF term in the TALYS calculation.    

\subsubsection{\label{ss:reactions}Evaluation of first-chance reaction cross sections}

We can clearly subdivide present evaluation work for the (n+$^{239}$Pu) cross sections in two parts: the lowest energy range below $B_n^{^{239}Pu}$, meaning below 5.65~MeV where second-chance reactions are limited to a small contribution in the fission channel (cf. Fig.~\ref{fig:Pgfi239*}) and, the higher energy range  up to third-chance fission. Following paragraphs are devoted to the common treatment of first-chance-only reactions as summarised in sections~\ref{ss:HFf}~and~\ref{ss:HFMC}. A glimpse of our evaluated work  has been proposed by  Fig.~\ref{fig:pu240cnxs}. It comes out immediately that  radiative decay (n,g) and compound nucleus elastic (n,n) cross sections can be neglected straight above a few MeV of neutron energy. \\

\begin{figure}[]
\center{\vspace{1.0cm}
\resizebox{0.99\columnwidth}{!}{
\includegraphics[height=5cm,angle=0]{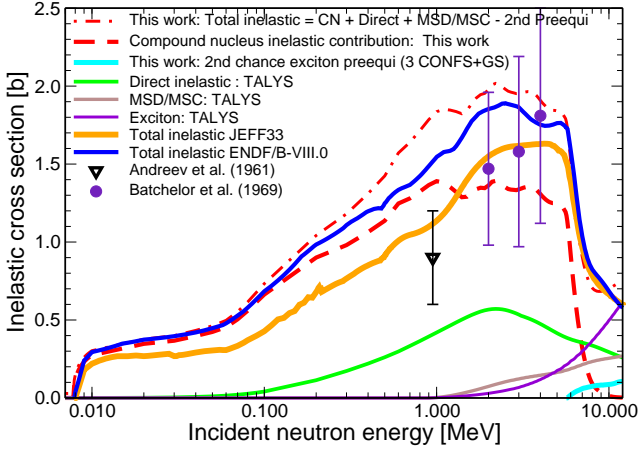}} }
\caption{\label{fig:totinel}(Color online) Present work scattering components involved in the (n+$^{239}$Pu) reaction that were summed   incoherently  to form the total inelastic cross section as function of neutron energy. Our inelastic total cross section is compared to the experimental data sets by Andreev~\cite{and:63} and, R.~Batchelor and K.~Wyld~\cite{bat:69},  and to the  ENDF/B-VIII.0~\cite{bro:18} and JEFF-3.3~\cite{plo:20} recommendations. Preequilibrium results depending upon the TALYS option selected, are shown as well. Final value of the total inelastic cross section is also related to the subtracted amount of preequilibrium component to the (n,2n)  second-chance cross section.}
\end{figure}

\paragraph{Inelastic cross section}  Between a few MeV and 5.65~MeV, the only  significant contributions to the total cross section are  the  total inelastic and  fission reactions with the former being approximated to a simple sum of  CN,  preequilibrium and  direct components. Over this medium-energy domain, the  sum relies essentially on an accurate calculation of the CN and direct components (Fig.~\ref{fig:totinel}). We are pretty confident in our calculation for  the former and the  precision of the latter is strongly correlated to any reasonable fit previously performed on measured total cross section data. Since our reconstructed total inelastic cross section is consistent with  the ENDF/B-VIII.0~\cite{bro:18} recommendation, there are arguments to think that the JEFF-3.3~\cite{plo:20} curve is underestimated over the range from the inelastic threshold up to the inelastic cusp (Fig.~\ref{fig:cusp}). Above 7~MeV, the exact magnitude of the total inelastic cross section is strongly connected to an exact prediction of the preequilibrium cross section that increases rapidly with energy if the exciton model is chosen. Latter model used in present work, did not offer the  small amount of total inelastic scattering suggested by  ENDF/B-VIII.0 or JEFF-3.3 around 12~MeV. Therefore an alternative calculation has been performed using the MSD/MSC model normalized on the value quoted in the evaluated files at 11~MeV that sounds to make reference~\cite{mcn:01}. This alternative is presented briefly in Appendix~\ref{unc} for completeness and for an estimation of the global error carried by present fitted model.\\

\begin{figure}[]
\center{\vspace{1.0cm}
\resizebox{0.99\columnwidth}{!}{
\includegraphics[height=5cm,angle=0]{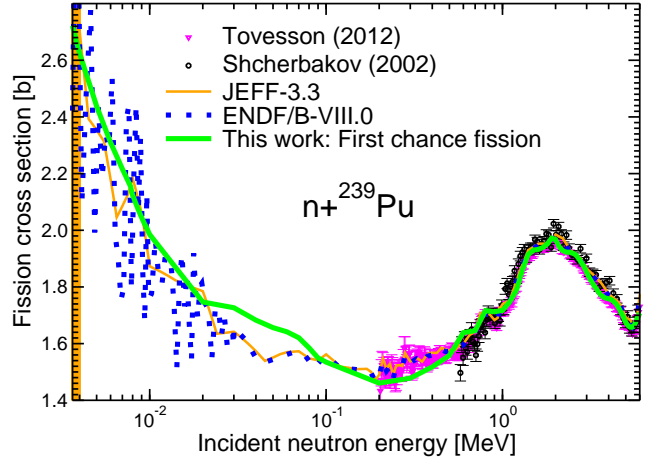}} }
\caption{\label{fig:fiss1}(Color online) $^{239}$Pu fission cross section profile as a function of neutron energy below $B_n^{^{239}Pu}$=5.65~MeV. Comparison of the present adjusted fission cross section with the evaluations (ENDF/B-VIII.0~\cite{bro:18} and JEFF-3.3~\cite{plo:20}) and the recent experimental data by Tovesson \etal~\cite{tov:10} and  Shcherbakov \etal~\cite{shc:02}.}
\end{figure}

\paragraph{Fission cross section}
A characteristic of present fission cross section evaluation work is the underlying theory based on a modification of the  Hauser-Feshbach statistical theory of nuclear reactions to treat the fission decay channel in the R-matrix formalism~\cite{lyn:73}. In particular the fluctuations of the fission decay widths due to the presence of intermediate structures in a second well of the overall fission barrier and thus to the coupling of class-I and class-II states, are simulated by Monte Carlo sampling of the underlying model parameter distributions. Use of standard HF equations (cf. Eq.~\ref{eq:hfhfwconcicef}) for treating the fission channel is not so much of an issue when fitting fission cross sections but it leads to less physical fission parameters. In particular the absence of the class-II state width fluctuation correction factor, the  $W_{II}$ factor in the Eq.~(\ref{eq:FXSdecoupled}) of magnitude close to 0.8 in the resonance region, is necessarily compensated by additional tuning of the barrier heights. Use of the  extensive fission cross section formalism (Eq.~(\ref{eq:FXSdecoupled})) is then recommended; formalism that is still more accurate in its Monte Carlo form (Eq.(~\ref{eq:xsmc}); see also Ref.~\cite{bou:13}) because it does not imply any decoupling hypothesis  in the calculation of the HF fission cross section. \\

\begin{figure}[]
\center{\vspace{1.0cm}
\resizebox{0.99\columnwidth}{!}{
\includegraphics[height=5cm,angle=0]{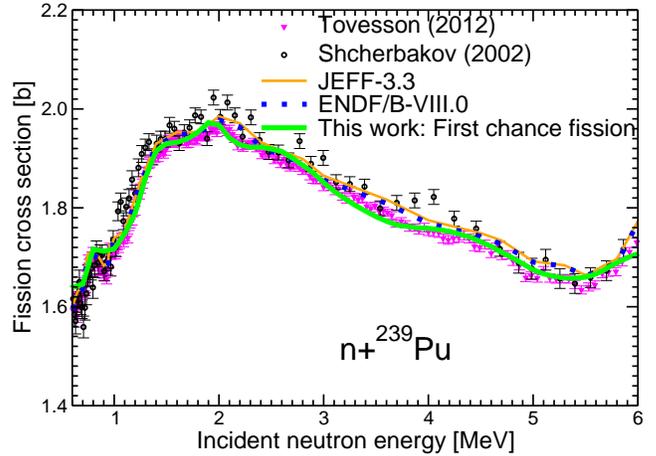}} }
\caption{\label{fig:fiss1b}(Color online) Same as Fig~\ref{fig:fiss1} but zoomed over the [0.5-6.0]~MeV energy range.}
\end{figure}

Present R-matrix formalism still falls into the statistical frame with actual average parameters obviously not designed to reproduce the medium-size structure commonly observed in the measured  fission cross section. Figure~\ref{fig:fiss1} shows our best reproduction of the measured $^{239}$Pu  neutron fission cross section as treated within the R-matrix formalism. It follows reasonably well the observed profile below 600~keV and quite well above with good agreement with from one side the ENDF/B-VIII.0~\cite{bro:18} and JEFF-3.3~\cite{plo:20} evaluated curves and, from the other side the recent experimental data by Tovesson \etal~\cite{tov:10} and  Shcherbakov \etal~\cite{shc:02}. The good agreement shown above 600~keV (Fig.~\ref{fig:fiss1b}) is crucial since it governs reciprocally the other sizeable exit channel; meaning the CN inelastic channel. In that sense it brings more confidence on the magnitude of the total inelastic cross section as calculated in present work. (Fig.~\ref{fig:totinel}; dot-dashed curve).\\ 

\subsubsection{Second-chance reactions}

\begin{figure}[]
\center{\vspace{1.0cm}
\resizebox{0.99\columnwidth}{!}{
\includegraphics[height=5cm,angle=0]{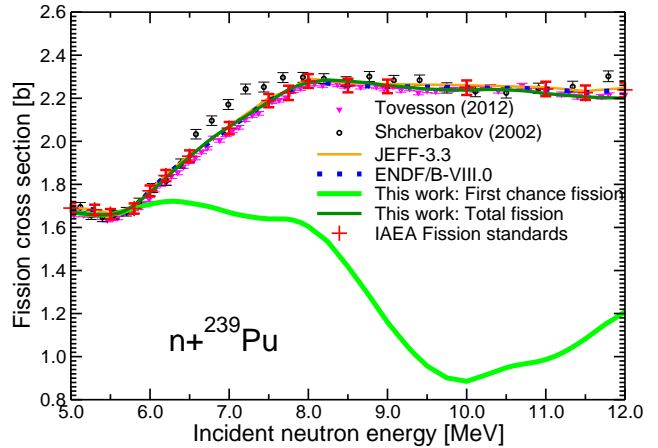}} }
\caption{\label{fig:fissHI}(Color online) Same as Fig~\ref{fig:fiss1} but above 5~MeV neutron energy. The fission cross section with uncertainties recommended by the IAEA neutron data standards committee~\cite{car:18} is drawn to give the best reference.}
\end{figure}

\begin{figure}[]
\center{\vspace{1.0cm}
\resizebox{0.99\columnwidth}{!}{
\includegraphics[height=5cm,angle=0]{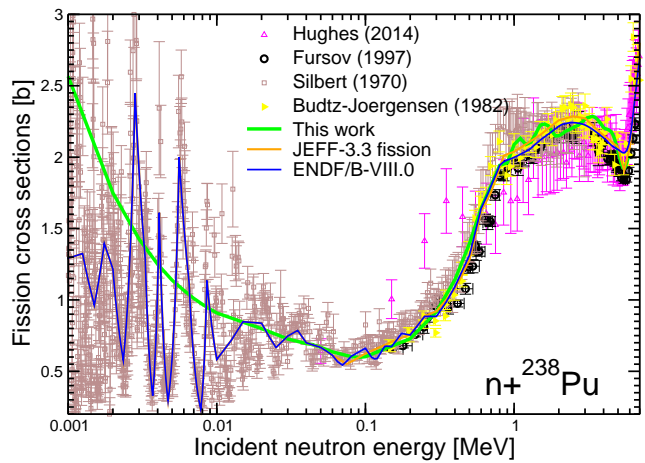}} }
\caption{\label{fig:fiss238}(Color online) $^{238}$Pu fission cross section profile as a function of neutron energy below second-chance fission. Present adjusted fission cross section is compared to the  ENDF/B-VIII.0~\cite{bro:18} and JEFF-3.3~\cite{plo:20} evaluations and, the experimental data by Silbert \etal~\cite{sil:73}, Budtz-Joergensen \etal~\cite{bud:82}, Fursov \etal~\cite{fur:97} and  Hughes \etal~\cite{hug:14}.}
\end{figure}

\begin{figure}[]
\center{\vspace{1.0cm}
\resizebox{0.99\columnwidth}{!}{
\includegraphics[height=5cm,angle=0]{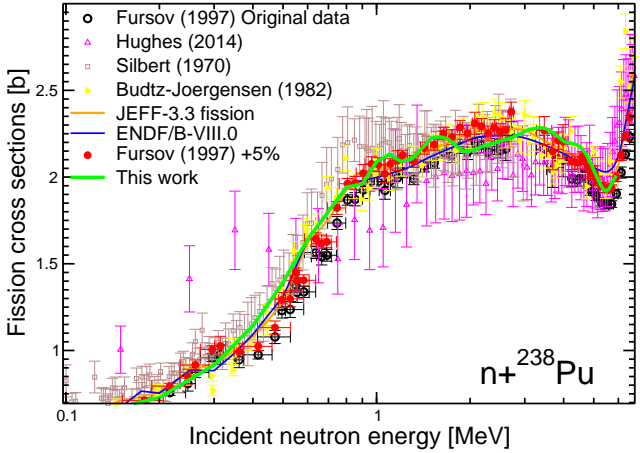}} }
\caption{\label{fig:fiss238z}(Color online) Same as Fig~\ref{fig:fiss238} but zoomed over the high-energy range. The experimental data trend carried by Silbert \etal~\cite{sil:73} and Budtz-Joergensen \etal~\cite{bud:82} suggests an increase of 5\% for the recent measurement by Fursov \etal~\cite{fur:97}.}
\end{figure}

At about 5~MeV neutron energy, the (n,n'f) second-chance fission channel opens (Fig.~\ref{fig:Pgfi239*}) and  soon above the (n,2n) reaction at $B_n^{^{239}Pu}$. This picture is fulfilled with competitive radiative deexcitation in the $^{239}$Pu  first-chance-residual nucleus. Second-chance reactions are fed mainly by the HF inelastic neutron  emission decay channel of the $^{240}$Pu compound nucleus. Therefore, again, comes the importance of the choice of the measured fission cross section that makes reference. This choice and the quality of  present fit on the fission cross section selected is emphasized by Fig.~\ref{fig:fissHI}. Present evaluated fission cross section follows the  recommended curve by the IAEA neutron data standards committee~\cite{car:18}. This recommendation is strong since the uncertainties assessed by the IAEA within the range of second-chance reactions are quite small and always close to 1.5\%. Making an alternative choice as the one addressed by the fission cross section measurement by Shcherbakov \etal~\cite{shc:02} would give another perspective on the (n,2n) evaluated cross section, especially within the range [6.5-8.0]~MeV (Fig.~\ref{fig:fissHI}). Consequences of latter alternative together with the choice of another preequilibrium model (as the MSD/MSC proposed by TALYS) is commented in the Appendix~\ref{unc}. Among the ingredients to support the best prediction for the $^{239}$Pu (n,2n) cross section is the accuracy of the calculated reaction decay probabilities for the $^{239}$Pu* (Fig.~\ref{fig:Pgfi239*}). The latter was triggered by the parameters evaluated during a prior fit of the $^{238}$Pu neutron cross sections (Fig.~\ref{fig:pu239cnxs}) and in particular over the high-energy range below second-chance fission. Figure~\ref{fig:fiss238} shows in detail our evaluated  neutron fission cross section according to the $^{238}$Pu target nucleus. Choice has been made to make our confidence in the data by Fursov \etal~\cite{fur:97} but with a renormalization of +5\% to reach agreement with the older data sets of Silbert \etal~\cite{sil:73} and Budtz-Joergensen \etal~\cite{bud:82} (Fig.~\ref{fig:fiss238z}).  The most recent experimental data set~\cite{hug:14} derived from surrogate measurements using the  surrogate reaction method~\cite{bou:20}, has been judged not accurate enough by comparison to the results from neutron 'spectroscopy'.\\

\begin{figure}[]
\center{\vspace{1.0cm}
\resizebox{0.99\columnwidth}{!}{
\includegraphics[height=5cm,angle=0]{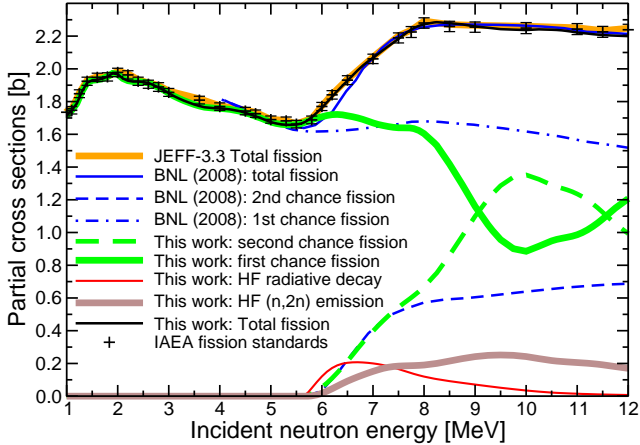}} }
\caption{\label{fig:overview}(Color online) Pattern of second-chance reactions and components as a function of neutron energy according to the (n+$^{239}$Pu)  system above 1~MeV. A comparison is proposed with the comprehensive study (labeled BNL) by Maslov \etal\cite{mas:08}. Present fitted total fission cross section follows the curve recommended by the IAEA neutron data standards committee~\cite{car:18}.}
\end{figure}

Finally we give an overview of the  calculated (n,n'$\gamma$), (n,n'f) and (n,2n) compound nucleus second-chance reaction cross sections according to the (n+$^{239}$Pu)  system above 1~MeV (Fig.~\ref{fig:overview}) with comparison to the comprehensive results obtained by Maslov \etal\cite{mas:08}. It reveals strong differences with~\cite{mas:08} in terms of first-chance and second-chance fission profiles above 7.5~MeV. {\it The actual pattern of the fission cross section components is strongly correlated to the fitting procedure taken that involves the choice of the experimental fission cross sections making reference, the model of preequilibrium selected and the accuracy of the optical model calculations performed at the beginning of the process}. Present results are also driven by the competition among first- and second-chance fissions and first-stage-only inelastic across the Hauser-Feshbach engine.   
%Differences encountered on first chance and second-chance fission profiles with the latter study are attributed to the present sequential procedure that involves prior parameter fits on the series of measured Pu isotopes cross section data involved and a final (n+$^{239}$Pu) HF calculation that involves careful subtraction of the neutron flux lost during the nucleus preequilibrium phase; meaning a strong correlation with the preequilibrium model selected. We must keep in mind that this is essentially the total fission cross section  magnitude calculated  and in competition thereof with the HF inelastic cross section, which is important for correct prediction of second-chance reactions HF neutron flux.

\subsection{\label{s:n2n} Total (n,2n) excitation function}

We now reach  the target set in the introduction, meaning an actual capability to predict with reasonable precision the total (n,2n)  cross section as a function of neutron energy that merges both CN and preequilibrium second-chance contributions. The latter depends obviously on the model of level density selected for the residual nucleus. According to the {\it exciton} preequilibrium model, a residual nucleus level density taken as the LD {\it after} compound nucleus emission (CNLD) is expected to return good results. However using a LD spin distribution based on {\it a few quasi-particle excitations} sounds to be more consistent with a non-equilibrated residual nucleus process. In present work, we have tested several types of LD for the $^{239}$Pu residual nucleus to estimate the  range of uncertainty covering present total (n,2n)  cross section prediction. The first  LD tested was assuming a preequilibrium phase restrained to excited quasi-particle states formed by the breaking of one pair of neutrons and one pair of protons in addition to the ground state excited one-neutron quasi-particle states;  ensemble of configurations noted \{0P-1N\},\{0P-3N\},\{2P-1N\}. The second one was involving a slightly more complicated preequilibrium phase before neutrons escape from the composite system by including the \{0P-5N\} configuration. Present  tests have been completed with the classic CNLD and an extreme case of state population represented by the total level density. As suggested before, full extend of present work must include testing the spin distribution of the residual nucleus provided by one of the most promising microscopic method; meaning QRPA-based  results as supplied by the Ref.~\cite{dup:17}. Since the QRPA results corresponding to the $^{239}$Pu residual nucleus were not available, we have selected the spin distribution obtained for the residual nucleus formed after the preequilibrium emission of one neutron in the neighboring (n+ $^{238}$U) reaction. Latter distribution, so far limited to the contribution of the  natural parity states, $\pi= (-1)^J$  (Ref.~\cite{dup:17}),  is expected to carry into our calculation the singular pattern commonly shown by QRPA results. Fig.~6 of Ref.~\cite{dup:17} shows a weak excitation energy dependence of the QRPA spin distribution but a sharp distribution centered around $J^\pi=3^-$. We recall that  present calculations assume valid for the calculation of the second-chance preequilibrium cross section component the second-chance neutron-emission probabilities as derived from the Hauser-Feshbach equations.\\ 

\begin{figure}[t]
\center{\vspace{0.cm}
\resizebox{0.95\columnwidth}{!}{
\includegraphics[height=5cm,angle=0]{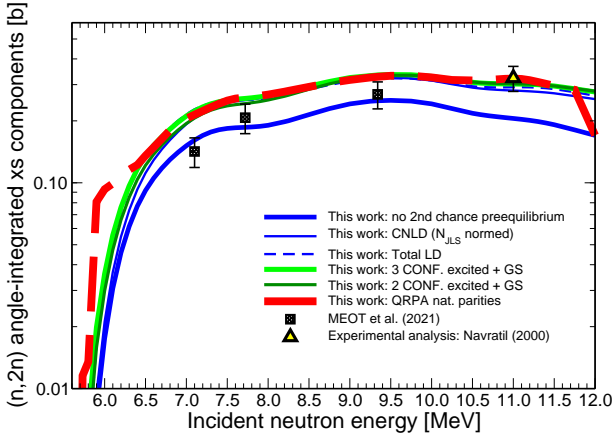}}}
\caption{\label{fig:allDist}(Color online) Total $^{239}$Pu(n,2n)  cross section calculated as a function of both neutron energy and  type of level density selected for the residual nucleus. This graphic displays the CN (n,2n) cross section component (bottom curve) as well as the sum of Hauser-Feshbach and  preequilibrium second-chance contributions. The latter involves successively the (\{0P-1N\}, \{0P-3N\}, \{2P-1N\}) and the (\{0P-1N\}, \{0P-3N\}, \{2P-1N\}, \{0P-5N\}) combined configurations, a full equilibrated residual nucleus (CNLD) and the total level density. Results using the QRPA spin distribution of Ref.~\cite{dup:17}, are included as reference. Experimental data by Navratil and McNabb~\cite{nav:00}, and M\'eot \etal~\cite{meo:21} are drawn as well.}% as the modeled calculation by Blann.~\cite{bla:84}}}
\end{figure}

\begin{figure}[t]
\center{\vspace{0.cm}
\resizebox{0.95\columnwidth}{!}{
\includegraphics[height=5cm,angle=0]{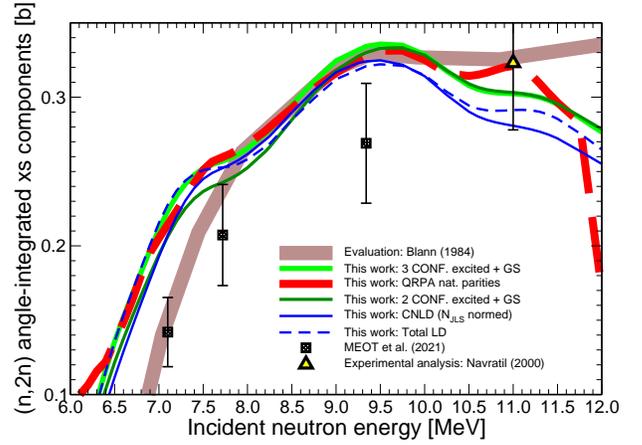}}}
\caption{(Color online) Same as Fig~\ref{fig:allDist} but using linear y-scale. The figure now includes the reference evaluation (solid thick brown curve) made by Blann and White~\cite{bla:84} in 1984.}
\label{fig:allDistz}
\end{figure}

Present various calculations performed for the total (n,2n)  cross section are displayed in Fig.~\ref{fig:allDist} including a comparison with the single Hauser-Feshbach contribution. We promptly  observe that the  HF component represents the largest part of the total (n,2n)  cross section but also carries part of the wavy shape of the final curve depending upon the model of residual nucleus level density selected (as demonstrated in the Appendix~\ref{uncLD}). This wavy shape reproduces well the data points of M\'eot \etal~\cite{meo:21} which are, in magnitude, lower than the sum of the HF and  preequilibrium second-chance  components. Fig.~\ref{fig:allDist} is appropriate to enlighten the strong difference over the threshold energy region of the excitation function in between  the QRPA-based curve and the others. Additional differences among the various level-density-based curves are visible in Fig.~\ref{fig:allDistz}. The later also includes the {\it reference} evaluation made by Blann and White~\cite{bla:84} in 1984. This evaluation was based on a Weisskopf-Ewing model calculation with a pre-equilibrium model and experimentally-based fission probabilities. We observe in Fig.~\ref{fig:allDistz} an excellent agreement between Blann and White~\cite{bla:84} and our calculations over the [8-10]~MeV range. The differences lie within the threshold of the reaction and above 10~MeV, depending upon the level density model selected. It is important to remark that present QRPA-based curve meets the Navratil and McNabb~\cite{nav:00} data point at $E_n = 11$~MeV which is considered as a true reference since quoting the authors {\it 'It is solidly based on nuclear data and it includes a measure of the  $^{239}$Pu(n,2n) cross section independent of the other direct measurements'}. Figure~\ref{fig:allDistz}  shows also that the QPVRLD-based curves remain within the uncertainties of the point at $E_n = 11$~MeV, noting however that the agreement is better for the non-equilibrated-LD-based calculations. Finally, it enlightens the good consistency among present calculations over a wide energy domain whatever the LD model is; in practice from 6.5 up to 9.5~MeV. From the above considerations, we will quote as {\it in-house preequilibrium reference} the calculation that assumes the (\{0P-1N\}, \{0P-3N\}, \{2P-1N\}, \{0P-5N\}) combined configurations for the LD of the residual nucleus since the related curve demonstrates the best agreement with the QRPA-based simulation (including the change of slope at 7.5~MeV).\\ 

\begin{figure}[t]
\center{\vspace{0.cm}
\resizebox{0.95\columnwidth}{!}{
\includegraphics[height=5cm,angle=0]{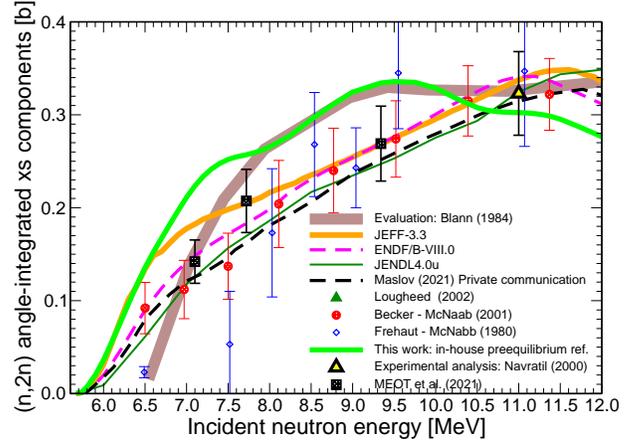}}}
\caption{(Color online) Present total $^{239}$Pu(n,2n) cross section calculation is compared to the JEFF-3.3~\cite{plo:20}, ENDF/B-VIII.0~\cite{bro:18} and JENDL4.0u~\cite{jendl4u}  evaluations and to the most confident experimental database. Experimental values plotted are extracted from the 2001 evaluation report by McNabb \etal~\cite{mcn:01} that includes slight downwards renormalization  of Fr\'ehaut \etal~data. For  completeness, a recent proposal by Maslov~\cite{mas:21a} is shown as well.}
\label{fig:n2nNewArt}
\end{figure}

\begin{figure}[t]
\center{\vspace{0.cm}
\resizebox{0.95\columnwidth}{!}{
\includegraphics[height=5cm,angle=0]{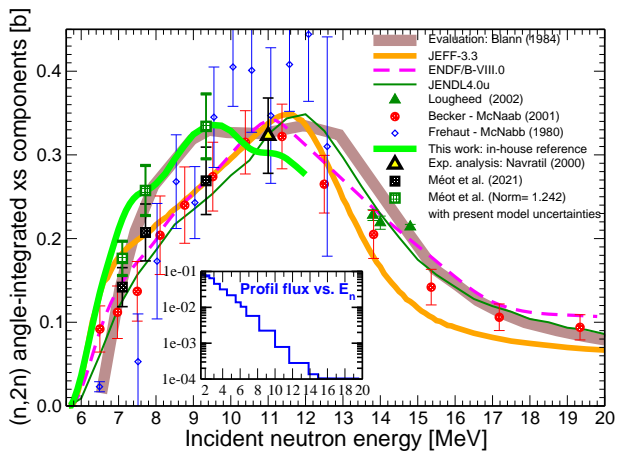}}}
\caption{(Color online) Same as Fig~\ref{fig:n2nNewArt} but expanded over the whole neutron energy spectrum. Above graphic draws the experimental data set of M\'eot \etal~after renormalization together with our estimation of the present model uncertainties. The inset reports the neutron flux in arbitrary unit as measured in the PROFIL experiment~\cite{tom:08} according to the incident neutron energy [MeV]}
\label{fig:n2nProposal}
\end{figure}

Figure~\ref{fig:n2nNewArt} benchmarks our in-house reference with the most confident experimental database together with major evaluated curves. It reveals promptly that our reference curve follows the initial slope of the JEFF-3.3 data~\cite{plo:20} over the threshold but then quickly deviates significantly upwards with a maximum at much lower energy  than the peak predicted in the previous studies (at about 9.5~MeV compared to 11.6~MeV).  It happens that the present (n,2n) maximum is related to the minimum of first chance fission as visible on Fig.~\ref{fig:overview}. Various attempts to shift this maximum towards higher energies have failed. Although it makes sense to remain cautious in our prediction over the [10-12]~MeV energy range since this domain is close to third-chance reactions, only mimicked in present calculation, we remain confident in our calculation because of the safe agreement with Navratil and McNabb~\cite{nav:00} at $E_n=11$~MeV. Finally, an extrapolation of the present calculation suggests a (n,2n) high-energy trend truly compatible with the experimental results published by Becker \etal~\cite{bec:02} and Loughheed~\cite{lou:02} (Fig.~\ref{fig:n2nProposal}).\\

In their recent paper, M\'eot \etal~\cite{meo:21} brought to our attention the results of the PROFIL integral experiment~\cite{tom:08}, carried out in the PHENIX fast neutron spectrum reactor, which returned some (n,2n) feedback different from both the theoretical calculations and the microscopic measurements previously cited. Using the JEFF-3.1 evaluation data set library, the authors of Ref.~\cite{tom:08} have obtained a value of 0.793(34) for the ratio of the calculation to the experiment according to the $^{239}$Pu(n,2n)  cross section integrated over the PHENIX fast neutron spectrum. Since within the [6-8]~MeV energy region (Fig.~\ref{fig:n2nArt}) the JEFF-3.1 $^{239}$Pu(n,2n)  cross section~\cite{tom:08} is even 6\% higher than the JEFF-3.3 cross section~\cite{plo:20}, the disagreement between the integral feedback and the current JEFF-3.3 data library is expected to be similar. Unfortunately the recent microscopic measurement by M\'eot \etal~\cite{meo:21} at respectively 7.1, 7.72 and 9.34~MeV did not clarify this issue since the new data points are within the uncertainties addressed by other differential measurements (except Fr\'ehaut \etal~\cite{fre:80}) (Fig.~\ref{fig:n2nProposal}). By enhancing the left-tail of the total (n,2n)  cross section, present calculation makes a bridge between the integral feedback~\cite{tom:08} and the theoretical expectation since above 1~MeV neutron energy, the neutron flux measured for the PROFIL experiment decreases rapidly (Fig.~\ref{fig:n2nProposal}, inset). As a response to the PROFIL experiment issue we calculated, using the new (n,2n) cross section prediction complemented above 12~MeV by the JEFF-3.1 data file (Fig.~\ref{fig:n2nArt}), the modified (n,2n) resonance integral to be compared with the JEFF-3.1 original value. The resonance integral is defined as, \\     
\begin{eqnarray}
I_{n,2n}= \int_{E_{threshold}}^{20~MeV}  dE\mbox{ }\sigma_{n,2n}(E)\Phi_{n}(E)  \mbox{ , }
\label{eq:I2n}
\end{eqnarray}
with $E$, the neutron energy and $\Phi_{n}$ the neutron flux as measured in the PHENIX experiment (Fig.~\ref{fig:n2nProposal}, inset). The  increase of $I_{n,2n}$ brought by present work, relatively to the JEFF-3.1 evaluation, is estimated at {\bf +7.1\%}. This result is consistent with  the general trend extracted from the study~\cite{tom:08}, based on  integral validation of the JEFF-3.1 data file and where, quoting Tommasi and Nogu\`ere~ {\it 'The integral (n,2n) reactions checked were underestimated by 7\% to 30\%}', depending upon the selected target actinide.  
%-- TABLE -- RESIDUAL DISCRETE
\begin{table}[ht]\center{
\caption{\label{tab:lowLying} Low-lying discrete levels sequence up to 1.1~MeV as input to present calculation for the $^{238}$Pu compound nucleus. Data supplied are extracted from RIPL~\cite{cap:09} and complemented when  information is missing by QPVR simulations. Whenever relevant, comments are added according to ENSDF~\cite{dim:20}.}
\footnotesize{\resizebox{1.0\columnwidth}{!}{\begin{tabular}{lrcl}
\hline\hline
Energy & $I$ & $\pi$ & Comments \\
(MeV) & ($\hbar$)&  &  \\
\hline
0.0000			&   0	& +  & 	ground-state (gs) band 0   \\
0.0441	&   2	& +  &  gs rotational  band 0 - ripl 1 u   \\
0.14595	&   4	& +  &  gs rotational  band 0 - ripl 1 u   \\    
0.3037	&	  6 & +  &  gs rotational  band 0 - ripl 1 u   \\
0.5136	&	  8 & +  &  gs rotational  band 0 - ripl 1 u   \\
0.6051  &  1  &-  &  mass asym.   band 1 - ripl 2 u   \\
0.6614  &  3  &-  &  mass asym.   rotational  band 1 - ripl 2 u   \\
0.7632  &   5 &- &  mass asym.   rotational  band 1 - ripl 2 u   \\
0.7735  &  10 & +  &  gs rotational  band 0 - ripl 1 u   \\
0.9096  &  7  &-  &  Not in RIPL: inserted for completeness; mass asym. rot. band 1; confirmed ENSDF-2014 at 0.9116    \\	
0.9415  &  0  & +  &  Beta band 2	 - ripl 3 u   \\
0.9628  &  1  &-  & 	2 quasi-particles (qp) band	 - ripl 4 u as $J^\pi= 2^-$ not $1^-$   \\
0.9682  &  2  &-  & 	ENSDF file offers no ch.  - ripl 2 u   \\			
0.9831  &  2  & +  & 	Beta rotational band 2  - ripl 5 u   \\		
0.9628  &  1  &-  &  Not in RIPL: bending band 3?    \\		 
0.9855  &  2  &-  &  ripl 3 u   \\
1.0190   &  3  &-  &  ENSDF2014 no sign given (suggested rotational  - ripl 1 g as Jpi= 3+ not 3-   \\
1.0180   &  3  &-  &  Not in RIPL: inserted for completeness	   \\
1.0283  &  2  & +  &  ENSDF file offers no comments (same as band 5?) - ripl 3 u    \\
1.0650  &  2  &-  &  inserted for completeness		   \\
1.0700   &  3  & +  &  ripl 2 u   \\
1.0830   &  4  &-  &  ENSDF suggests 2 qp state - ripl 2 u   \\		
1.1010   &  9 &-  & Not in RIPL: mass asym.   band 1 (inserted for completeness); confirmed ENSDF-2014 at 1.1024 \\
\hline\hline
\end{tabular}}}}
\end{table}	

\section{\label{ccl}conclusion} 

Present work has been launched in response to the recent  measurement performed by M\'eot \etal~\cite{meo:21} of the $^{239}$Pu$(n,2n)^{238}$Pu reaction cross section based on the recoil method for counting the $^{238}$Pu nuclei. Latter measurement brings some clarification on the exact profile of the left tail of the (n,2n) excitation function. Present study comes after a series of hard work and differential measurements on that topic, well represented by the US task force~\cite{mcn:01} or the efforts made in the frame of the JEFF project~\cite{for:97,rom:16,plo:20} to resolve the issue. {\it By the present study, we do not intend to end this debate but rather to bring some hope for an earlier conclusion}.\\  

In this work we have followed the footsteps of Romain \etal~\cite{rom:16} using the 'full model' methodology that ensures consistency between all types of cross sections over the Pu isotopes family although present implemetation could not cope accurately with third-chance reactions, justifying some precaution according to the results within the [10-12]~MeV neutron incident energy range. Present analysis was made possible by chaining the TALYS-ECIS06~\cite{kon:12} system of codes with an upgraded version of the AVXSF-LNG computer program~\cite{bou:13,bou:18} that is now able to deal with second-chance reactions using its decay-reaction-probabilities feature~\cite{bou:19,bou:20}. A key issue to achieve the objective set in preamble, was a prior simultaneous analysis~\cite{bou:13} of the neutron-induced average cross-sections of the plutonium isotopes family from 236 to 244 for neutron energies within the [1~keV - 5.5~MeV] range. Present 2021 database is pretty much the same of this prior study for the lower energy range where substantial effort has been put in building a suitable database of nuclear structure parameters. By using TALYS-ECIS06 as an input to the AVXSF-LNG Hauser-Feshbach engine, a possible weakness of the historical AVXSF-LNG methodology based on even-odd $S_l$ neutron-energy-dependent strength functions, was corrected to achieve calculations at above 1~MeV neutron energy with a better accuracy. \\

Present diligent study brings some confidence in the results obtained as illustrated in Fig.~\ref{fig:n2nProposal} (green wavy solid curve) at least up to 10~MeV although present curve extrapolation to higher energies remains in good agreement with all modern age evaluations. Left tail of the present (n,2n) cross section calculation exhibits a marked wavy shape that is confirmed by the experimental data points released by M\'eot \etal~\cite{meo:21}. This wavy shape could have been expected because of the even-even character of the $^{238}$Pu  second-stage-residual nucleus. By anticipation, a careful description of the low-lying  levels of $^{238}$Pu  based mainly on RIPL information~\cite{cap:09} has been build up to 1.11~MeV excitation energy (Tab.~\ref{tab:lowLying}). On the ground of the arguments developed along this paper, it sounds reasonable to state that current evaluated files under-estimate  the $^{239}$Pu(n,2n) cross section below 10~MeV; under-estimation of the order of 7.1\% relatively to the JEFF-3.1 evaluation. This result is supported by the analysis of the PROFIL integral experiment~\cite{tom:08} suggesting an increase of the (n,2n) resonance integral up to 21\%. Since the recent measurement by M\'eot \etal~\cite{meo:21} is  normalized somehow arbitrarily on the data point of Becker~\cite{ber:01,bec:02} at 9.52~MeV which carries an  uncertainty of $\pm$ 15\%, there is definitively some room for advising a correction of normalisation for the  experimental values recently published~\cite{meo:21}. In present calculation of the (n,2n) cross section, the value obtained at 9.34~MeV is of 334~mb in comparison to 269~mb claimed by M\'eot \etal~This suggests a renormalization factor of {\bf 1.24} with a maximum uncertainty on the present calculation of {\bf $\pm$  11.6\%}; estimated from the difference between our reference calculation and the alternative path as described in Appendix~\ref{unc}. Figure~\ref{fig:n2nProposal} draws the experimental data set of M\'eot \etal~after renormalization together with our fitted model uncertainties. We verify that present calculation wavy shape is in reasonable agreement with the renormalized data set, even with the data point at 7.1~MeV.

\appendix
\section{\label{connect}$\sigma_{n}^{CN}$ reconstruction in AVXSF-LNG}
The neutron-induced compound nucleus formation cross section for given  ($J,\pi$) couple is conventionally defined as :
\begin{eqnarray}
\sigma_{n}^{CN}(E_n,J^{\pi}) = \pi \lambdabar^2 g_{_{J,I}} \sum _{s={|I-\frac{1}{2}|}}^ {{|I+\frac{1}{2}|}} \sum _{l={|J-s|}}^ {|J+s|}  T_n^{J^{\pi_{(ls)}} }(E_n)\mbox{
}
\label{eq:NC}
\end{eqnarray}
in which $g_{{J,I}}$ is the spin  factor $(2J+1)/(2(2I+1))$ and  $T_n^{J^{\pi_{(ls)}}}$ the neutron entrance transmission coefficients using {\it L-j coupling} scheme. In the absence of EW transformation, $T$ is more precisely called {\it generalised} transmission coefficient~\cite{kaw:16} and defined in {\it the direct cross-section-eliminated space} that assumes diagonal channels. A channel $c \equiv (l, s)$ is defined by its relative angular momentum $l$ and spin $s$ and where the summations obey the usual conservation rules. $I$ and $i$ characterize the intrinsic spins of the target nucleus and projectile respectively.\\

\begin{figure}[t]
\center{\vspace{0.7cm}
\resizebox{0.99\columnwidth}{!}{
\includegraphics[height=5cm,angle=0]{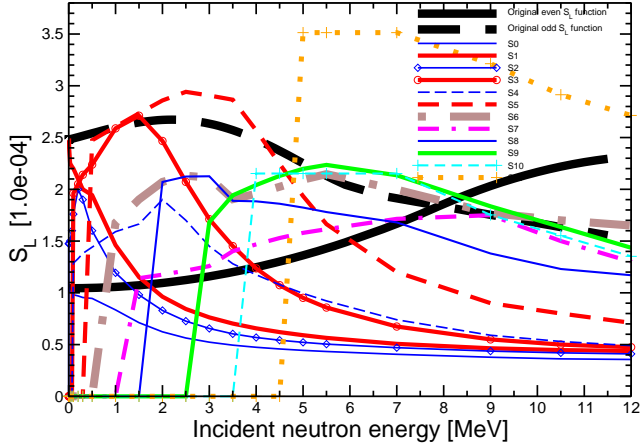}} }
\caption{\label{fig:pu239SL}(Color online) Historical even (thick solid black line) and odd (thick dash black line) l-waves neutron strength functions according to the (n+$^{239}$Pu) reaction as a function of neutron energy by comparison to the $S_l$ data set produced by the ECIS-06 code. Latter set is now used as input to the  AVXSF-LNG compound nucleus calculations.}
\end{figure}

In  our code,  computing of neutron channel transmission coefficients  follows the general form derived by Moldauer~\cite{mol:67}
\begin{eqnarray}
T_n^{J^{\pi_{(ls)}}}=1-exp\left(-2\pi S_l  \right),
\label{eq:tn}
\end{eqnarray}
where $S_l$ is the energy dependent neutron strength function for given relative orbital momentum $l$. Literature on heavy nuclides supplies accurate low energy (asymptotic)  values of $S_l$  for $s$- and $p$-wave neutron channels extracted from resolved resonance region analyses and average cross section fits below $300$ keV. According to the Pu isotopes family,  good estimates of $S_l$ are $1.044\times10^{-4}$ and $1.48\times10^{-4}$ respectively for both waves. These asymptotic values are mostly within the uncertainties addressed in associated literature ($\pm10\%$ and $\pm27\%$ at best respectively for $s$- and $p$-waves). The historical AVXSF-LNG treatment of strength functions was based on an empirical rule~\cite{fro:01} assuming close values for even l-waves and reciprocally for odd l-waves. This ad hoc method, with l-wave penetrability threshold and long-range energy dependence corrections, used in study~\cite{bou:13} has been substituted in present work by  a refined  prediction of the $S_l$ strength functions using the TALYS-ECIS06 nuclear reaction system of codes. Present work $S_l$ data set according to the (n+$^{239}$Pu) reaction is illustrated in Fig.~(\ref{fig:pu239SL}) with comparison to the historical even-odd $S_l$ neutron-energy-dependent strength functions. Nonetheless, the implementation of Eq.~\ref{eq:tn} still implies grouping partial strength functions prior to the resolution of the HF equations  such that,  
\begin{eqnarray}
S_l =   \frac{1}{(2l+1)}\sum _{ J={|s-l|}}^{|s+l|} g_{_{J,I}} \nu(l,J) S(l,J) \mbox{ ,}
\label{eq:sl}
\end{eqnarray}
where $\nu(l,J)$ and $S(l,J)$  are respectively the spin multiplicity according to each $l$ value and the partial neutron strength functions as a function of $l$ and $J$; all of them supplied by TALYS. Once the total CN formation cross section, based on the generalized optical model transmission coefficients with correction for preequilibrium effects, is reconstructed in AVXSF-LNG, the equilibrated nucleus is 'allowed' to decay among the many open exit reaction channels. 

\section{\label{uncLD} Sensitivity to the model of residual nucleus level density}
\begin{figure}[t]
\center{\vspace{1.0cm}
\resizebox{0.99\columnwidth}{!}{
\includegraphics[height=5cm,angle=0]{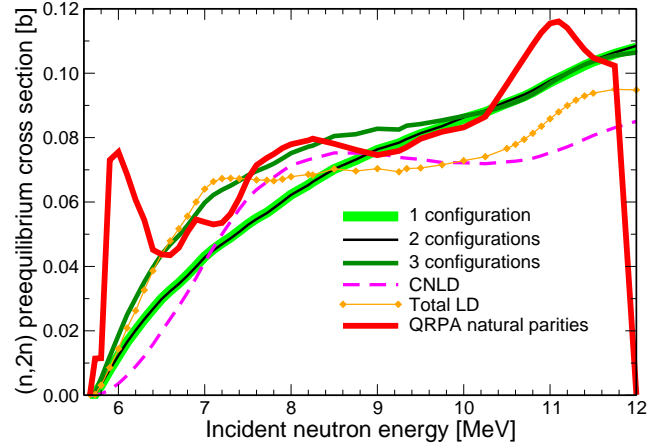}} }
\caption{\label{fig:n2npreequi}(Color online) (n,2n) preequilibrium cross section component as a function of neutron energy with dependence on the model of residual nucleus LD selected. Preequilibrium cross sections presented rely respectively one one (\{0P-3N\}), two (\{0P-3N\}, \{2P-1N\}) and three  (\{0P-3N\}, \{2P-1N\}, \{0P-5N\}) combined quasi-particle state configurations in supplement to the excited one-neutron ground state configuration (\{0P-1N\}). Those can be compared with the modeled cross sections relying on the CNLD, the total LD and finally the QRPA natural parity states LD of Ref.~\cite{dup:17} (Fig.~6) according to the $^{239}$Pu residual nucleus.}
\end{figure}

Figure~\ref{fig:n2npreequi} shows the dependence of  the (n,2n) preequilibrium cross section to the model of residual nucleus LD selected. We study the impact of an increase of the number of nucleon-pair breaking in the $^{239}$Pu non-equilibrated residual nucleus until a compound nucleus LD is reached. The extreme case of the total level density is presented as well as the most singular pattern introduced by a QRPA-type of LD. We understand that the choice made on the type of residual nucleus level density is critical to assess the actual shape of the total (n,2n) cross section.
 
\section{\label{unc} Sensitivity to alternative options}
\begin{figure}[h]
\center{\vspace{1.0cm}
\resizebox{0.99\columnwidth}{!}{
\includegraphics[height=5cm,angle=0]{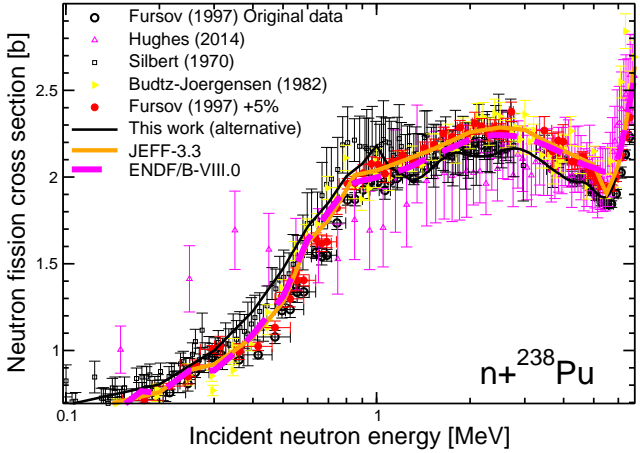}} }
\caption{\label{fig:238PuxsfAlt}(Color online) $^{238}$Pu fission cross section profile as a function of neutron energy above 1~MeV. Comparison of the present adjusted fission cross section with both the  ENDF/B-VIII.0~\cite{bro:18} and JEFF-3.3~\cite{plo:20} evaluations and, the experimental data by Silbert \etal~\cite{sil:73}, Budtz-Joergensen \etal~\cite{bud:82}, Fursov \etal~\cite{fur:97} (original and renormalized (+5\%) data) and  Hughes \etal~\cite{hug:14}.}
\end{figure}

We have made a list in Section~\ref{ss:general} of the most sensitive items in the calculation of the total (n,2n)  cross section. This paragraph is willing by taking alternative choices in the calculation route, to show the most extreme result we can get while keeping the same approach on the reaction process. This alternative calculation is well portrayed by the next series of graphics. Figure~\ref{fig:238PuxsfAlt} shows the screening selection in regards to the experimental $^{238}$Pu  neutron-induced fission cross section (black solid curve) that will make reference during the nuclear parameters adjustment prior to the second-chance fission-decay probability calculation. On the contrary to the choice made in the main text, we keep confidence in the measurement by Fursov \etal~\cite{fur:97} and we do not apply the +5\% renormalization factor. The immediate effect is a lowering of the second-chance fission probability and reciprocally an increase of first-chance fission to preserve the experimental level of the total fission cross section.\\ 

The second alternative step we take is the selection of the other type of preequilibrium, differing in shape and magnitude, to be subtracted from the reaction cross section; meaning the MSD/MSC model of TALYS in place of the exciton model. We emphasize that the MSD/MSC preequilibrium curve (the thickest curve in Fig.~\ref{fig:preequi}) was nevertheless renormalized upwards (+39\%) such that at 11~MeV  the sum of direct inelastic and preequilibrium be close to the total inelastic cross section addressed by JEFF-3.3. Despite this enhancement, the 11~MeV preequilibrium value remains 37\% lower than the value returned by the  exciton model. The direct consequence lies in a reduced preequilibrium flux subtraction (Eq.~(\ref{eq:totcn})) balanced by a strong lowering of the fitted  outer fission barrier level density in the $^{240}$Pu nucleus. When no renormalization is made on the MSD/MSC preequilibrium cross section, unreasonable lowering of the fitted level density is encountered, questioning present  magnitude returned by the MSD/MSC preequilibrium TALYS (default) calculation.\\  
\begin{figure}[t]
\center{\vspace{1.0cm}
\resizebox{0.99\columnwidth}{!}{
\includegraphics[height=5cm,angle=0]{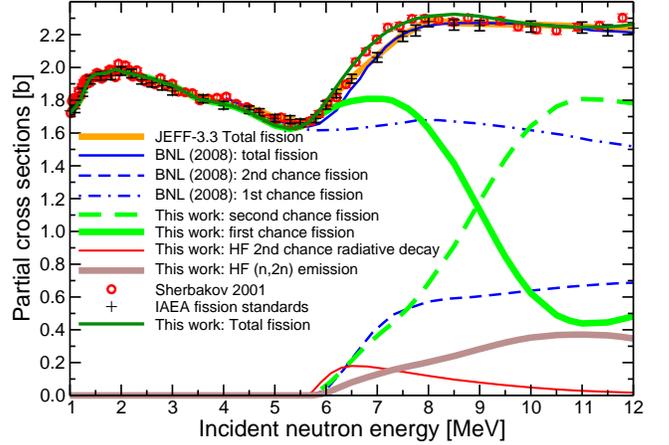}} }
\caption{\label{fig:overviewAlt}(Color online) Profile comparison of second-chance reactions and fission components as a function of neutron energy according to the (n+$^{239}$Pu)  system above 1~MeV. The above total fission cross section adjustment follows the experimental data of  Shcherbakov \etal~\cite{shc:02} in contrast to the recommended curve by the IAEA neutron data standards committee~\cite{car:18}.}
\end{figure}

\begin{figure}[]
\center{\vspace{1.0cm}
\resizebox{0.99\columnwidth}{!}{
\includegraphics[height=5cm,angle=0]{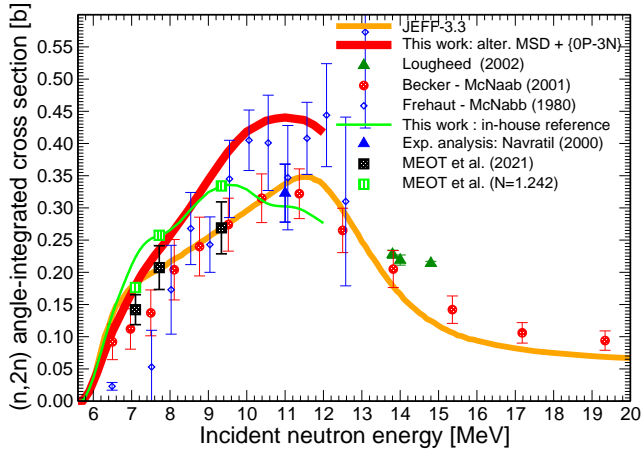}} }
\caption{\label{fig:n2nNewArtAlt}(Color online) This graphic shows the alternative calculation to the {\it in-house reference}, as described in the main text, with comparison to the current evaluations (JEFF-3.3~\cite{plo:20}, ENDF/B-VIII.0~\cite{bro:18}, JENDL4.0u~\cite{jendl4u}) and to the most confident experimental database for the total $^{239}$Pu(n,2n) reaction  cross section. Experimental values plotted are extracted from the 2001 evaluation report by McNabb \etal~\cite{mcn:01} that recommends slight downwards renormalization  of Fr\'ehaut \etal~data.}
\end{figure}

Third, we pick out the $^{239}$Pu neutron fission cross section as measured by Shcherbakov \etal~\cite{shc:02} in contrast to the IAEA neutron data standards committee recommendation~\cite{car:18} (Fig.~\ref{fig:overviewAlt}; red open circles). This boosts first-chance fission over the range [6.5-8.0]~MeV where the second-chance fission probability is still low as it is demonstrated by the comparison of Figures~\ref{fig:overviewAlt} and~\ref{fig:overview}. Last Figure~\ref{fig:n2nNewArtAlt} presents the $^{239}$Pu total (n,2n)  cross section (red thickest curve) resulting of these alternative options (keeping the original $^{238}$Pu neutron fission data of Fursov \etal, subtracting  a MSD/MSC preequilibrium-type cross section  and selecting the $^{239}$Pu fission data measured by Shcherbakov \etal). We conclude this alternative path with the selection of the lowest excited quasi-particle states configuration (\{0P-1N\},\{0P-3N\}) according to the preequilibrium phase in the residual nucleus (report also to  Fig.~\ref{fig:n2npreequi}).\\

By comparison to our {\it in-house reference}, we observe that the alternative curve is enhanced from 9.5~MeV  because of the excess of neutron flux feeding CN second-chance reactions (Fig.~\ref{fig:n2nNewArtAlt}). This trend, amplified as neutron energy increases, follows the lower slope gradient of the MSD/MSC model relatively to the {\it exciton} model. We note that in both alternative calculations,  a general trend to increase the total (n,2n)  cross section by comparison to JEFF-3.3 is verified in any case.   

%!!!---  April 2021  Talys preequilibrium calculation for n + target 239Pu renormalized at 11 MeV N=1.3914
%!!! Norm details Total inelastic evaluated at 11 MeV (on JEFF33 and BVIII.0) = 0.62784 b
%!!! sum of direct_rot5 + preequi4 @ 11 MeV = 0.528748
%suggests an increase of 5\% of the recent direct kinematics measurement by Fursov \etal~\cite{fur:97}. 

\acknowledgments
{\footnotesize One of the authors (O.B.) expresses his deep gratitude to Eric J. Lynn from LANL/T-2 for numerous fruitful discussions on physics related to the theory of nuclear reactions. O.B. is also indebted to P.~M\"oller (Scientific Computing and Graphics, Honolulu) for supplying with kindness full single-particle orbital database according to the actinides and to A.~K\"oning (IAEA) for complementary information about the total inelastic process. Special acknowledgement to G.~Nogu\`ere  and J.Tommasi (CEA/LEPh) for valuable feedback. Present work is dedicated to Dr. Eric Fort (1935-2017) who studied with passion the (n,2n) process at CEA/Cadarache.}

%P. MÂller  Scientific Computing and Graphics, P.O. Box 75009, Honolulu, HI 96836-0009, USA

\end{document}